\documentclass{article}

\usepackage[utf8]{inputenc}
\usepackage[T1]{fontenc}

\usepackage{geometry}
\usepackage{color}

\usepackage{lmodern}
\usepackage[tracking=true,kerning=true,final]{microtype}

\usepackage[unicode=true,pdfusetitle, bookmarks=true,bookmarksnumbered=false,bookmarksopen=false,
breaklinks=false,hidelinks,colorlinks=true]{hyperref}
\hypersetup{colorlinks=true, linkcolor=blue,  anchorcolor=blue, citecolor=blue,
filecolor=blue, menucolor=blue, urlcolor=blue}
\hypersetup{pdfauthor={},pdftitle={}}

\usepackage{caption}   
\usepackage{subcaption}
\usepackage{color}
\usepackage{graphicx}
\graphicspath{{../}{../figs/}{../pix/}{./}{./figs/}{./pix/}}

\usepackage{amsmath}
\usepackage{amsfonts}
\usepackage{bm}
\usepackage{mathtools}
\usepackage{amsthm}

\theoremstyle{definition}

\DeclareMathOperator*{\argmax}{arg\,max}
\DeclareMathOperator*{\argmin}{arg\,min}

\newcommand\numberthis{\addtocounter{equation}{1}\tag{\theequation}}

\title{Probabilistic characterization of the effect of transient stochastic loads on the fatigue-crack nucleation
time}

\author{Stephen Guth and Themistoklis P. Sapsis
\thanks{Corresponding author: \href{mailto:sapsis@mit.edu}{sapsis@mit.edu},
Tel: (617) 324-7508, Fax: (617) 253-8689%
}\\
Department of Mechanical Engineering,
\\ Massachusetts Institute of Technology, \\
77 Massachusetts Ave., Cambridge, MA 02139}
\date{\today}

\usepackage{amsmath}
\usepackage{amssymb}

\usepackage[normalem]{ulem}

\begin{document}

 \maketitle\ 
\begin{abstract}
The rainflow counting algorithm for material fatigue is both simple to implement and extraordinarily successful for predicting material failure times.  However, it neglects memory effects and time-ordering dependence, and therefore runs into difficulties dealing with highly intermittent or transient stochastic loads with heavy tailed distributions.  Such loads appear frequently in  a wide range of applications in ocean and mechanical engineering, such as wind turbines and offshore structures.  In this work we employ the  Serebrinsky-Ortiz cohesive envelope model for material fatigue to characterize the effects of load intermittency on the
fatigue-crack nucleation time. We first formulate efficient numerical integration schemes, which allow for the direct characterization of the fatigue life in terms of any given load time-series. Subsequently, we consider the case of stochastic intermittent loads with given statistical characteristics. To overcome the need for expensive Monte-Carlo simulations, we formulate the fatigue life as an up-crossing problem of the coherent envelope. Assuming statistical independence for the large intermittent spikes and using probabilistic arguments we derive closed expressions for the up-crossing properties of the coherent envelope and obtain analytical approximations for the probability
mass function of the failure time. The analytical expressions are derived directly in terms of the probability density function of the load, as well as the coherent envelope. We examine the accuracy of the analytical approximations and compare the predicted failure time with the standard rainflow algorithm for various loads. Finally, we use the analytical expressions to examine the robustness of the derived probability distribution for the failure time with respect to the coherent envelope geometrical properties.

  
\end{abstract}
\newpage
\tableofcontents

\section{Introduction}

Modern engineering applications increasingly rely on enormously capital intensive structures, which are placed in extreme conditions and subject to extreme loads that vary significantly throughout the structure expected lifetime.  Failure costs are astronomical, including forgone profits, legal penalties, tort payouts, and reputation damage \cite{lee18}.  Minimizing lifetime costs require safe-life engineering and a conservative assessment of failure probabilities.  Unfortunately, while material fatigue is a major contributor to failure, non destructively measuring fatigue is both difficult and expensive \cite{zhang01}.

For many classes of structures, fatigue loads have a stochastic character with transient features that cannot be captured through a statistically stationary consideration. Examples include loads in wind turbines due to control and wind gusts, transitions between chaotic and regular responses in oil risers, and slamming loads in ship motions \cite{khan07, hu16, wolfsteiner17}.  For such applications, traditional frequency domain approaches have difficulty predicting the fatigue effects of intermittent loading, as those are inherently connected to time-ordering and therefore cannot be captured by the spectral content of the load.

An alternative approach for the prediction of fatigue-crack nucleation relies on hysteretic cohesive-law models \cite{serebrinsky05, arias06}. The appeal of this type of model of fatigue-crack nucleation is its applicability to arbitrary geometries, loading conditions, and loading histories, including random load cycling with transient spikes. However, quantifying the failure time under stochastic loading typically requires the generation of a large number of loading time histories and the computational solution of the hysteretic cohesive-law models for each of these loading scenarios. This Monte-Carlo scheme is associated with an enormous computational cost.  Alternative approaches include statistical linearization \cite{spanos12, murata14, chernyshov16}, hierarchical modeling \cite{zio08, mohamad16}, structured sampling methods \cite{ohlson03}, and optimal experimental design \cite{forrester08, malkomes16, ky16, huan13, jiang17}.  While there is a plethora of uncertainty quantification methods that can be potentially applied for the statistical characterization of the failure time there have been very few research endeavors on this direction. Moreover, one has to pay particular attention as very few uncertainty quantification methods can take into account the effect of transient intermittent events. 

In this work, we first develop an efficient time-marching scheme for the fatigue model developed by Serebrinksy and Ortiz \cite{serebrinsky05, arias06} with take into account the time-ordering effects.  Based on this model we then derive analytical approximations for the probability mass function (pmf) of failure time in terms of the load probability density function (pdf) and the coherent envelope.  We demonstrate the developed ideas in several examples involving intermittent loads. We also compare with standard fatigue life estimation methods and discuss their performance in various loading scenarios.
Finally, we examine the robustness of the derived pmf with respect to geometrical properties of the coherent envelope.

\section{The Serebrinsky-Ortiz (SO) model}
\label{sec:analytical-so-model}

We consider a single material element with one dimensional loading.  The applied load is a random process given by $\sigma(t)$, and the corresponding displacement, $\delta(t),$ depends on the element constitutive relation.  Further, this constitutive relation depends on the fatigue state of the material; after some number of loading/unloading cycles, $N_{f}$, the material stiffness will degrade and eventually the material will fail.  Our aim is in the relationship between the load $\sigma(t)$ and the failure time $N_f$.

For the characterization of
fatigue-crack initiation we consider the  hysteretic cohesive-law model
by 
Serebrinsky and Ortiz \cite{serebrinsky05, arias06}. 
In this model the fatigue-crack nucleation problem is formulated as a
first-upcrossing problem of the opening displacement-load curve: $(\delta(t),\sigma(t))$
with a critical boundary: \textit{the cohesive envelope}. 

We begin with a brief description of the cohesive model and explain in what
sense it can be used for the quantification of fatigue-crack
nucleation. A cohesive-law, i.e. the critical boundary is expressed as a
curve in the $\delta,\sigma$ plain. When the curve $(\delta(t),\sigma(t))$
meets the descending branch of the monotonic cohesive envelope the material
interface loses stability and we have the nucleation of fatigue-crack.
The cohesive envelope is typically described by the relation \begin{equation}
\sigma=\mathcal{F}(\delta)\triangleq e\sigma_c\frac{\delta}{\delta_c}e^{-\frac{\delta}{\delta_c}},
\label{cohevn}
\end{equation}
where $\sigma_c, \delta_c$ are constants that characterize the material.

The next step is to characterize the evolution of the opening displacement
$\delta(t)$ in terms of a given loading history $\sigma(t)$. We will employ
a simple phenomenological model \cite{serebrinsky05, arias06}\begin{equation}
\dot \sigma=\begin{cases}K^-\dot \delta, \ \ if \ \ \dot \delta<0,   \\
K^+\dot \delta, \ \ if \ \ \dot \delta>0,  \\
\end{cases}
\label{sigmaeq}
\end{equation}
where $K^+$ and $K^-$ are the loading and unloading incremental stiffnesses,
respectively. 

For simplicity we assume that unloading always takes place towards the origin
and is determined by the unloading point. This condition fully defines $K^-$.
By contrast, the loading stiffness, $K^+$, is assumed to evolve in accordance
with the kinetic relation \begin{equation}
\dot K^+=\begin{cases}(K^+-K^-)\frac{\dot \delta}{\delta_a}, \ \ if \ \ \dot
\delta<0,   \\
-K^+\frac{\dot \delta}{\delta_a}, \ \ if \ \ \dot \delta>0,  \\
\end{cases},
\label{kappaeq}
\end{equation}
where the fatigue endurance length, $\delta_a$, is a characteristic opening
displacement. The initial loading stiffness is given by $K_0^+=\frac{\sigma_0^+}{\delta^-}$,
where  $\sigma_0^+$ is the first peak of the load function, and $\delta^-$
is the corresponding value of  the ascending part of the cohesive envelope. Assuming that a typical initial load is much smaller than $\sigma_c$ we can express $K_0^+$ in terms of the coherent envelope as
\begin{align}\label{init_s}
K^+_0=\lim_{\delta \rightarrow 0} \frac{\mathcal F(\delta)}{\delta}.
\end{align}
If a subsequent load peak results in intersection with the ascending
part of the cohesive envelope, i.e. for $\delta \le \delta_c$ then a new
initial
loading stiffness is defined and it evolves according to equation (\ref{kappaeq}). This typically happens during extreme loading events and it is a mechanism that results in strong variability on the fatigue-lifetime of the material. By simple integration one can identify the evolution of $\delta(t)$
and $K^+(t)$ and estimate when the $(\delta(t),\sigma(t))$ curve will eventually intersect the
descending part of the cohesive envelope, when we have fatigue-nucleation. 

The presented model allows to take
into account not only the number and amplitude of the cycles but also their
specific sequence. In this sense it can be employed to study the effect of
noise but also high-frequency intermittent loads acting on the structure.

\subsection{Time-discretization of the SO model} 
\begin{figure}[t]
\begin{center}
\includegraphics[width=0.45\linewidth]{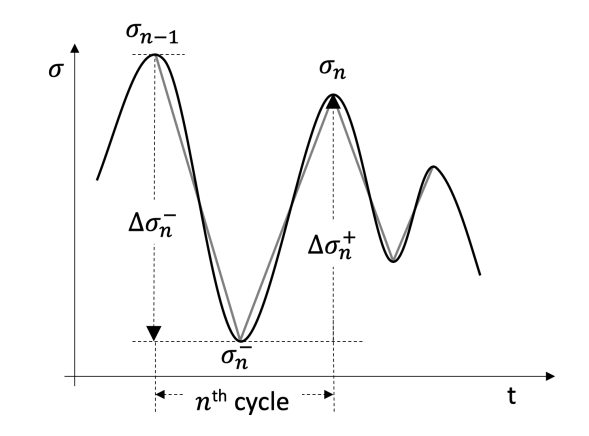} 
\end{center}
\caption{Piece-wise linear approximation of the loading function $\sigma(t)$
separates the loading and unloading part of the cycle and leads to an analytical
approximation. }
\label{fig2}
\end{figure}

We assume that the loading function $\sigma(t) $ is a continuous random function
with positive values. If the loading function becomes negative the opening
displacement vanishes and the loading stiffness remains constant until the
next positive load.
Therefore, we assume a positive load and we approximate $\sigma(t)$ with
a piece-wise linear function between local maxima/minima, i.e. \begin{align*}
\dot \sigma(t)&=\frac{\Delta  \sigma^-_n}{\Delta t^{-}_n}, \ \ \  t\in [\tau_{n-1},\tau_{n-1}+\Delta
t^{-}_n],\\
\dot \sigma(t)&=\frac{\Delta  \sigma^+_n}{\Delta t^{+}_n}, \ \ \  t\in [\tau_{n-1}+\Delta
t^{-}_n,\tau_{n-1}+\Delta
t^{-}_n+\Delta
t^{+}_n],
\end{align*}
where $n$ is the number of cycle, $\Delta  \sigma^-_n$ ($\Delta \sigma_n^+$)
is the negative (positive) increment of the cycle, i.e. $\Delta \sigma_n^-<0$
($\Delta \sigma_n^+>0$), and $\Delta t_n^-$ ($\Delta t_n^+$) is the corresponding
duration (Figure \ref{fig2}). The initial time of the $n^{th}$ cycle is denoted
as $\tau_{n-1}$, the initial load and opening displacement (both local maxima)
are denoted as,  $\sigma_{n-1}$ and $\delta_{n-1}$, respectively, while $K_{n-1}^+$
  represents the loading stiffness.
In addition, $\delta_n^-$  denotes the local minimum of the
opening displacement during the $n^{th}$ cycle, i.e. when $\sigma=\sigma_n^-$.

This approximation allows us to derive a solution in the form
of an iterative map. We  first express the evolution of the opening displacement
during the descending part of the cycle, i.e. after the increment, $\Delta
\sigma^-_{n}$. In this case,

\begin{equation}
K^-_{n}=\frac{\sigma_{n-1}}{\delta_{n-1}},
\label{load_min}
\end{equation}
 Therefore, by direct integration we have the evolution of the opening displacement
during unloading 
 \begin{align}
\delta^-_{n}&=\delta_{n-1}+\frac{\Delta \sigma^-_{n}}{K^-_{n}}=\delta_{n-1}\left(
1+\frac{\Delta \sigma^-_{n}}{\sigma_{n-1}} \right).
\label{disp_min}
\end{align}
Moreover, during unloading we have, by combining equations
(\ref{sigmaeq}) and (\ref{kappaeq}) (note that $K^+<K^-$):  \begin{align*}
\frac{dK^+}{K^{-}-K^+}=-\frac{\dot \sigma}{K^-\delta_a}.
\end{align*}
Integrating by parts we obtain,\begin{equation}
\frac{K^-_n-K^+_{n,u}}{K^-_{n}-{K^+_{n-1}}} =\exp \left ( \frac{\Delta \sigma^-_n}{\delta_aK^-_n} \right),
\label{unload_pos_stif}
\end{equation}
or equivalently,
\begin{equation}
K^+_{n,u}= K^-_n-\exp \left(\frac{\Delta \sigma^-_n}{\delta_aK^-_n} \right)(K^-_{n}-{K^+_{n-1}}),
\label{eeqxc}
\end{equation}
where $K^+_{n,u}$ is the loading stiffness right after the end of the negative
(unloading) increment of the cycle. Next, we compute the opening displacement
after the positive increment of the cycle. We first compute the evolution
of the loading stiffness, $K_n^+$ during loading.
We combine equations
(\ref{sigmaeq}) and (\ref{kappaeq}), to obtain \begin{equation}
\dot K^+=-\frac{\dot \sigma}{\delta_a},
\end{equation}
which is\ integrated to obtain the value of $K^+_{n}$ during the loading
part of the $n^{th}$ cycle:

 \begin{equation}
K^+_{n}(\sigma)=K^+_{n,u}-\frac{\sigma}{\delta_a}, \ \ \sigma \in [0,\Delta
\sigma_n^+]
\label{load_stif_d}
\end{equation}
At the end of the cycle we will have\begin{equation}
K^+_{n}=K^+_{n,u}-\frac{\Delta
\sigma^+_n}{\delta_a}.
\label{loadingp}
\end{equation}
The initial loading stiffness is given by $K_0^+=\frac{\sigma_0^+}{\delta^-}$,
where  $\sigma_0^+$ is the first peak of the load function, and $\delta^-$
is defined through the ascending part of the cohesive envelope,
as the solution of the equation\begin{equation}
\sigma_0^+=e\sigma_c\frac{\delta^-}{\delta_c}e^{-\frac{\delta^-}{\delta_c}},
\ \ \ \delta^- \le \delta_c.
\label{asc_coh}
\end{equation}
If a subsequent load peak $\sigma^+_n$ results in intersection with the ascending
part of the cohesive envelope, i.e. for $\delta \le \delta_c$ then a new
initial
loading stiffness is defined and it evolves according to equation (\ref{kappaeq}).

\begin{figure}[bht]
\begin{center}
\includegraphics[width=0.45\linewidth]{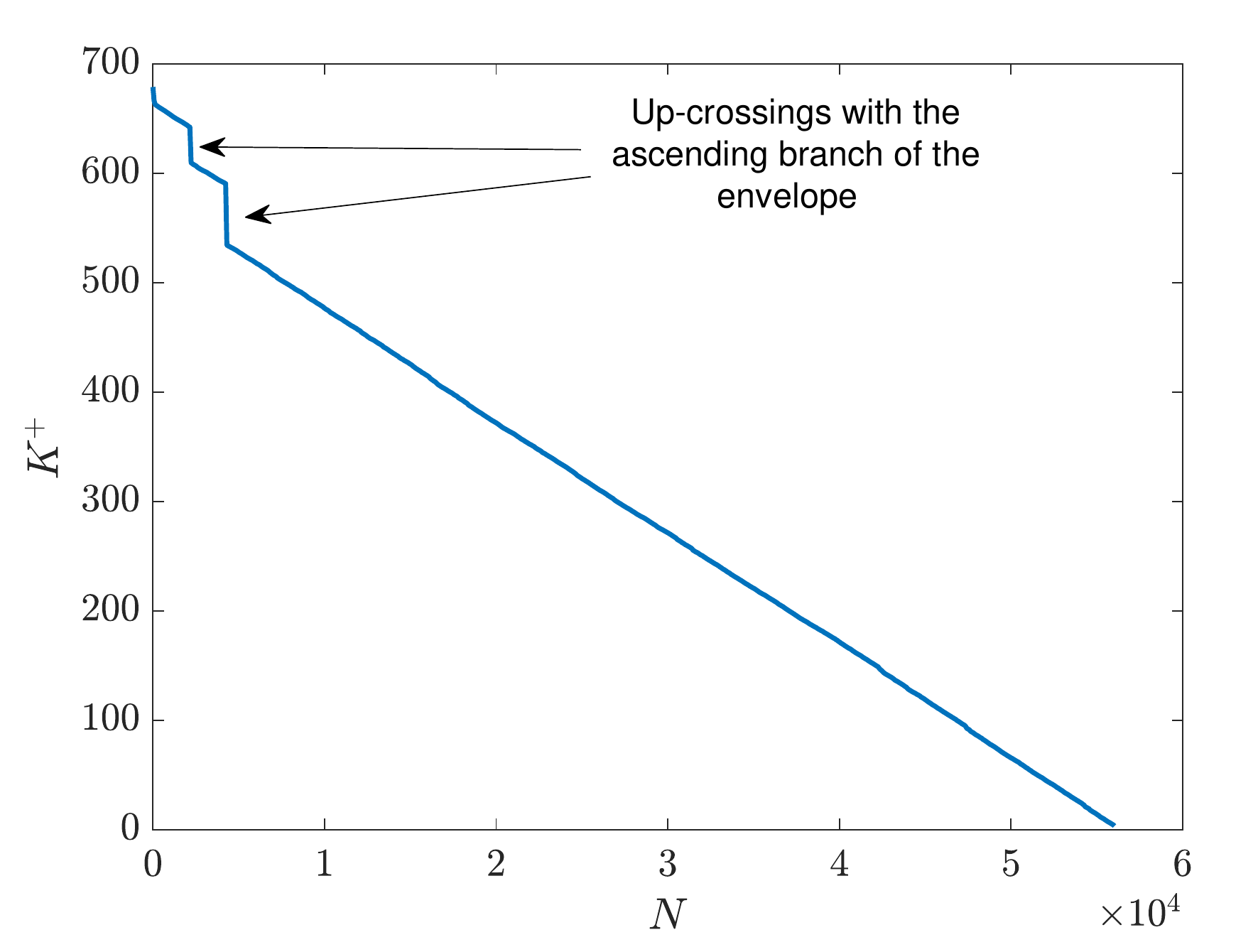} 
\end{center}
\caption{Schematic of the hysteretic cohesive-law and corresponding definition
of fatigue-crack nucleation. The  numbers close to the cohesive envelope
curve indicate the number of loading cycles. The solid lines show loading
and the dashed lines represent unloading.}
\label{fig3}
\end{figure}

\begin{figure}[]
\centering
\includegraphics[width=0.55\linewidth] {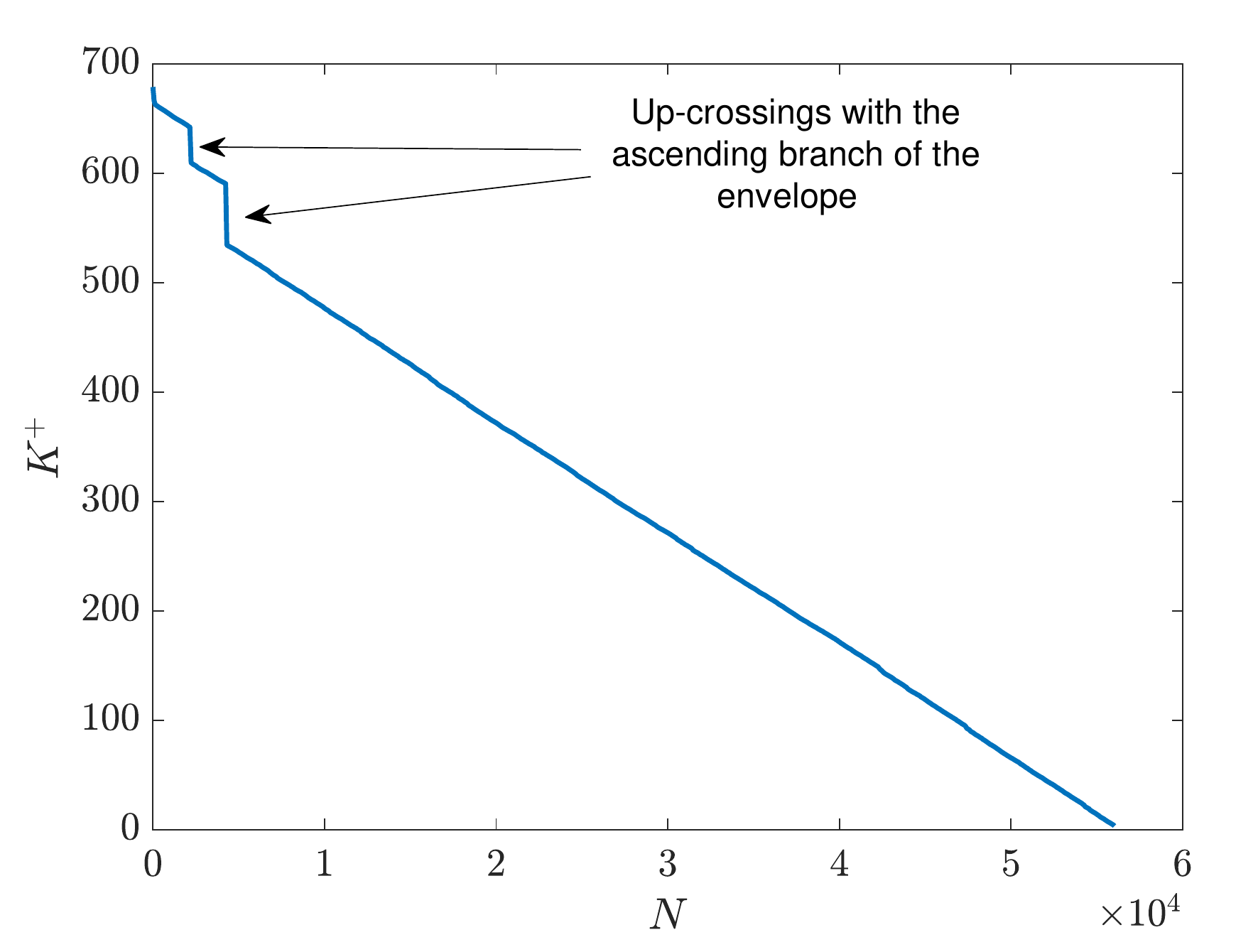}
\caption{Sample time evolution of $K^+$ for the SO model. Discontinuities indicate up-crossings of the ascending branch of the coherent envelope.}
\label{fig:K-skip}
\end{figure}

We utilize expression (\ref{load_stif_d}) into eq. (\ref{sigmaeq}): \begin{equation}
d\delta=\frac{d\sigma}{K^+_{n,u}-\frac{\sigma}{\delta_a}}.
\end{equation}
By integration we have,\begin{equation}
 \delta_n-\delta_n^-=-\delta_a \log\left(1-\frac{\Delta\sigma_n^+}{\delta_a
K_{n,u}^+}\right),
\label{delta_evol}
\end{equation}
which is the local maximum of the opening displacement history, $\delta(t)$,
at the end of the $n^{th}$ cycle. Crossing of this quantity with the descending part of the cohesive
envelope
(\ref{cohevn}) is an indication of fatigue-crack nucleation. Equations
(\ref{disp_min}), (\ref{unload_pos_stif}), (\ref{load_stif_d}) and (\ref{delta_evol})
provide
a piece-wise linear approximation of the opening displacement evolution, $\delta_n$. A graphical
summary of the described scheme is given in Figure \ref{fig3}.
 Based on the equation of the cohesive envelope (\ref{cohevn}), we have
material failure for the minimum number,  $N_f$, for which we have an up-crossing of the descending part of the cohesive envelope:  \begin{equation}
\sigma_{N_{f}}^+ \ge e\sigma_c\frac{\delta_{N_{f}}^+}{\delta_c}e^{-\frac{\delta_{N_{f}}^+}{\delta_c}} \ \ \ and \ \ \  \delta^+_{N_f}>\delta_c.
\end{equation}
A sample of the evolution of the stiffness, $K^+_{n}$, is shown in Figure \ref{fig:K-skip}. We observe that the evolution is typically linear except of several discrete jumps. These jumps are associated with crossings of the ascending part of the coherent envelope due to intermittent loading events.
In the next section we formulate an approximation scheme that takes into account the almost linear evolution of the stiffness away from extreme events and significantly accelerates the computation.

\subsection{Simulation of a loading time series using the probabilistic decomposition-synthesis method}


Here we formulate an approximation scheme for the failure-time under arbitrary loading time series. We first decompose the loading signal into segments associated with extreme events and regular loading events. For segments of the loading time-series where there is no up-crossing of the coherent envelope, we will show that the evolution of $K^+$ can be linearly approximated:
\begin{equation}
\label{eq:linear-k}
        K^+_n - K^+_{n - 1} \approx \Delta K.
\end{equation}
where $\Delta K$ has a constant value that we will estimate later.
The remaining segments of the loading time series are associated with the discontinuous jumps in Figure \ref{fig:K-skip}, corresponding to the intersections of the $(\sigma, \delta)$ curve with the ascending part of the coherent envelope.  This breakdown of an intermittent process into a linear region and an extreme region parallels the probabilistic decomposition-synthesis framework developed in \cite{mohamad15, joo18}.

Let the sequence $\{\sigma^+_n\}$ be the discretization of $\sigma(t)$ such that $\sigma^+_n$ is the $n^{\mbox{th}}$ local maxima of $\sigma(t)$, and let $\hat{\sigma}$ be a fixed threshold.  This sequence may be broken into two sets:

\begin{itemize}
\item $S_Q = \{n : \sigma^+_n \leq \hat{\sigma} \}$  -- the linear (quiescent) region
\item $S_I = \{n : \sigma^+_n > \hat{\sigma} \}$ -- intermittent spikes
\end{itemize}

For points in the set $S_Q$, we will use the simplified update equation (\ref{eq:linear-k}).  For points in the set $S_I$, we will use the full SO update step as described in the previous section, regardless of whether the $(\sigma, \delta)$ curve actually crosses the coherent envelope.
Finally, for technical reasons we will remove the first few peaks ($n \leq N_{\mbox{init}} \approx 5$) from $S_Q$ and add them to $S_I$.  This helps to initialize the algorithm for the case when there are otherwise few or no early spikes in the set $S_I$.

\subsubsection{Estimation of the slope $\Delta K$}\label{slope}
We can directly estimate the slope $\Delta K$ in (\ref{eq:linear-k}) from the SO model and the input signal statistics.  Combining equations (\ref{eeqxc}) and (\ref{loadingp}) we have\ 

%
%

\begin{align}
        \Delta K^+_n & = \left(1 - \exp\left(\frac{\Delta\sigma^-_n}{\delta_a K^-_n}\right) \right) (K^-_n - K^+_{n-1}) - \frac{\Delta \sigma^+_n}{\delta_a}.
        \label{so-update-exp-form}
\end{align}
The parameter $\frac{\Delta\sigma^-_n}{\delta_a K^-_n}$, which has inverse dependence on the endurance length $\delta_a$, is small unless a very large spike happens for very small $K^-$, i.e., towards  the end of the material lifetime. Therefore, expanding the exponential to first order produces

\begin{equation}
        \Delta K^+_n = \frac{\Delta\sigma^-_n}{\delta_a} \left(\frac{K^+_{n-1}}{K^-_n} - 1\right) - \frac{\Delta \sigma^+_n}{\delta_a}.
\label{eq:delta-k-two-term}
\end{equation}
This first term on the right hand side of equation (\ref{eq:delta-k-two-term})\ depends on the descending increment of the input signal, but it also depends on the small quantity $\left(\frac{K^+_{n-1}}{K^-_n} - 1\right)$ and therefore may be neglected in the small-increment regime.  The second term is state independent, and is directly proportional to the size of the ascending increment of the input signal.
When these approximations are made, we may relate the increments of $K^+$ to the increments of $\sigma$ using the following expression:

\begin{equation}
\label{eq:delta-k-approx}
        \Delta K^+_n =- \frac{\Delta \sigma^+_n}{\delta_a}.
\end{equation}
Assuming known statistical characteristics for the loading process, while the condition $\sigma^+ < \hat{\sigma}$ holds, the pdf and expected mean for $\Delta K^+$ are given by
\begin{align*}
\numberthis
\label{eq:delta-k-pdf-1}
        f_{\Delta K^+} (k) & = \delta_a  f_{\Delta \sigma^+ | \sigma^+ < \hat{\sigma}}(-\delta_a k) \\
        \Delta K = \mathbb{E}[\Delta K^+] & = -\frac{\mathbb{E}[\Delta \sigma^+|\sigma^+<\hat \sigma]}{\delta_a}
        \numberthis
\label{eq:delta-k-pdf-2}
\end{align*}

\noindent where $\mathbb{E}[\cdot]$ is the expected value (ensemble mean) operator, and the conditional pdf is given by

\begin{equation}
        f_{\Delta \sigma^+ | \sigma^+ < \hat{\sigma}} (\Delta \sigma^+) = \frac{f_{\Delta \sigma^+} (\Delta \sigma^+) \times  \mathbb{I}(\sigma^+ < \hat{\sigma}) }{\mathbb{P}[\sigma^+ < \hat{\sigma}]},
\end{equation}

\noindent where $\mathbb{I}(\cdot)$ is the indicator function. We note that one can also estimate the mean value $\Delta K$ empirically through a direct simulation of a short segment of the loading history. Under this approximation and assuming  no up-crossing with the coherent envelope, the evolution of the stiffness can be approximated by: 
\begin{equation}
        K_{n}^+=K_0^+-n\Delta K,
\end{equation}
This is a valid assumption as long as the spread of values of $\Delta \sigma_n^+$ is not systematically large. In this case, the material life will extend until the material stiffness vanishes, i.e. 
\begin{equation}
        K_{N_f}^+=K_0^+-N_f\Delta K=0,\label{eq_noup}
\end{equation}
Below we consider some additional cases where analytical approximations can be obtained.



\paragraph{Narrow-banded Gaussian Process.} Suppose that $\sigma(t)$ is a narrow-banded zero-mean Gaussian random process with standard deviation $\varrho$.  Since the negative values do not matter for the SO model, we need to characterize only the positive peaks. The pdf for the peaks of the process is given by (see, for instance \cite{naess13}) 
\begin{align}
        f_{\sigma^+}(a) = \frac{a}{\varrho^2} \exp\left(-\frac{a^2}{2\varrho^2}\right), \qquad a \geq 0.
\end{align}



\paragraph{Narrow-banded Non-Gaussian Process.} Suppose instead that $\sigma(t)$ is narrow-banded but not Gaussian. If the process is zero-mean then we can rely again on the amplitude of the positive peaks. The pdf of the peaks can be found as,
\begin{equation}
\label{eq:rice-peak}
        f_{\sigma^+}(a) = \frac{-1}{v^+_{\sigma} (0)} \frac{d v^+_{\sigma} (a)}{da}, \qquad a \geq 0,
\end{equation}

\noindent where $v^+_\sigma(a)$ is the $a$-upcrossing rate, given by the Rice formula:
\begin{align}
\label{eq:rice-upcrossing}
        v^+_{\sigma} (a) & = \lim_{\Delta t \rightarrow 0} \frac{1}{\Delta t} \mathbb{E}[N^+(a, \Delta t)] 
         = \int_0^\infty \dot{\sigma} f_{\sigma\dot{\sigma}}(a, \dot{\sigma}) d\dot{\sigma}
\end{align}
For detailed derivations of these relations we refer to \cite{naess13}.

\paragraph{Full Increment Calculation.} Suppose that we relax the requirement that the distribution of peaks be an adequate substitute for the distribution of increments.  Following \cite{sorensen87}, we obtain an expression for the joint distribution of peak $a_1$, valley $a_2$, and gap $\tau$

\begin{equation}
\label{eq:sorensen-increment}
        f_{P, V}(a_1, a_2, \tau) = \frac{\int_{-\infty}^0 \int_0^{\infty} -\ddot{x}_1 \ddot{x}_2 f_{X_1 X_2 \dot{X}_1 \dot{X}_2 \ddot{X}_1 \ddot{X}_2} (a_1, a_2, 0, 0, \ddot{x}_1, \ddot{x}_2) d \ddot{x}_1 d \ddot{x}_2}{\int_{-\infty}^0 \int_0^{\infty} -\ddot{x}_1 \ddot{x}_2 f_{\dot{X}_1 \dot{X}_2 \ddot{X}_1 \ddot{X}_2} (0, 0, \ddot{x}_1, \ddot{x}_2) d \ddot{x}_1 d \ddot{x}_2},
\end{equation}

\noindent an upper bound on the distribution of $\tau$

\begin{equation}
        f_T(\tau) \leq \int_{-\infty}^0 \int_0^{\infty} -\ddot{x}_1 \ddot{x}_2 f_{\dot{X}_1 \dot{X}_2 \ddot{X}_1 \ddot{X}_2} (0, 0, \ddot{x}_1, \ddot{x}_2) d \ddot{x}_1 d \ddot{x}_2,
\end{equation}

\noindent and an expression for the distribution of increments $\Delta \sigma^+ =\sigma^+-\sigma^-$:

\begin{equation}
        f_{\Delta \sigma^+ }(e) = \int_0^{\infty} f_T(\tau) \int_{\infty}^{\infty} f_{P, V}(a, a-e, \tau) da d\tau.
\end{equation}
Equation (\ref{eq:sorensen-increment}) resembles a Rice-type equation, and it in turn may be simplified by assuming a Gaussian closure for the second derivatives.

\subsection{The rainflow counting algorithm}

\begin{figure}[]
\centering
\includegraphics[width=0.45\linewidth] {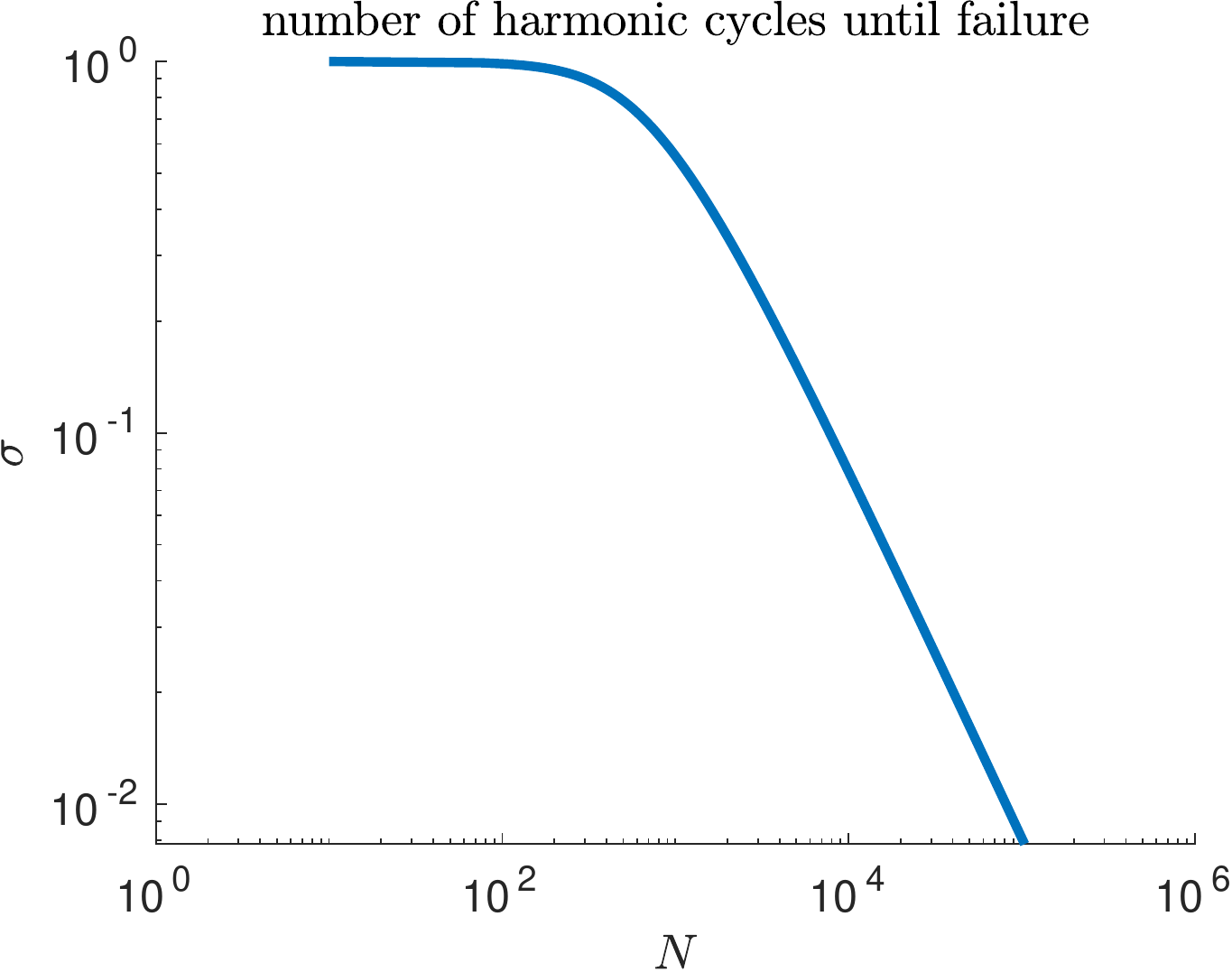}
\caption{Sample $SN$ plot for the coherent envelope in eq. (\ref{cohevn}) and the Serebrinksy-Ortiz fatigue model.}
\label{fig:so-sn-plot}
\end{figure}

The simplest method to characterize material failure properties is to apply a harmonic load with known amplitude, $S$, and count the number of cycles until material failure, $N$.  The relationship between $S$ and $N$ is a material characteristic; Figure \ref{fig:so-sn-plot} shows the $SN$ curve for the SO constitutive model.  Although realistic $SN$-curves are generally found to depend on the mean stress (compression versus tension) and loading (axial or torsional), simple cycle counting is found to work well for constant amplitude loading in the absence of intermittency \cite{schijve03}.
This frequency domain approach is easily adapted to non-harmonic signals by the Palmgren–Miner rule, given by

\begin{equation}
\label{eq:miner}
        \sum_i \frac{n_i}{N_{fi}} = C,
\end{equation}

\noindent which allows for calculation of equivalent fatigue by breaking the load signal into individual cycles, each of whose contributions is separately determined by reference to the $SN$-curve \cite{ciavarella18}. Based on the Palmgren–Miner rule the material fails when $C$ becomes greater than $1$.

The well known rainflow counting algorithm \cite{endo74} implements this rule by breaking a given signal into the corresponding set of increments.  The use of equation (\ref{eq:miner}) combined with a cycle-counting rule is standard in the literature on fatigue from random loads \cite{tatsis17, antoniou18, giagopoulos19}.
Note however that frequency domain methods such as rainflow counting completely ignores the time-ordering effects of the load history.

\subsection{A stochastic load model with intermittency}
To demonstrate the considered models we  employ a load produced
by a dynamical system exhibiting intermittent instabilities. This represents typical loads found in a wide range of engineering problems:\begin{align}
\begin{split}
\dot y+(\lambda -q_1z^2)y&=\nu_1 \dot W_y(t;\omega),\\
\ddot z + c\dot z + (k-q_2y)z  &=\nu_2 \dot W_z(t;\omega),\\
\sigma=Ez,
\label{driv1} 
\end{split}
\end{align}
where $\dot W_y, \dot W_z$ are white noise terms, $\lambda, k, c$ are constants, and $q_1, q_2$ are nonlinear coupling terms, and $E$ is the Young modulus. The model represents a structural mode, $z$, which interacts nonlinearly with another mode, $y$, creating intermittent energy transfers. It is a normal
form arising in a wide range of systems related to fluid-structure interaction
and mechanics, such as vortex induced vibrations, buckling, turbulent modes
and waves \cite{Sapsis21}. The $y-$mode represents the `reservoir' of energy (e.g. the axial
mode in buckling) while the $z-$mode represents the mechanical mode which
is excited by an additive noise as well as through intermittent energy transfers. The
nonlinear coupling terms are associated with intermittent energy transfers
from the $y-$mode to the $z-$mode. The displacement $z$ defines the evolution
of the stress $\sigma(t)$ that the material exhibits. 

\begin{figure}[ht]
\begin{center}
\includegraphics[trim=270 235 270 235,width=0.99\linewidth]{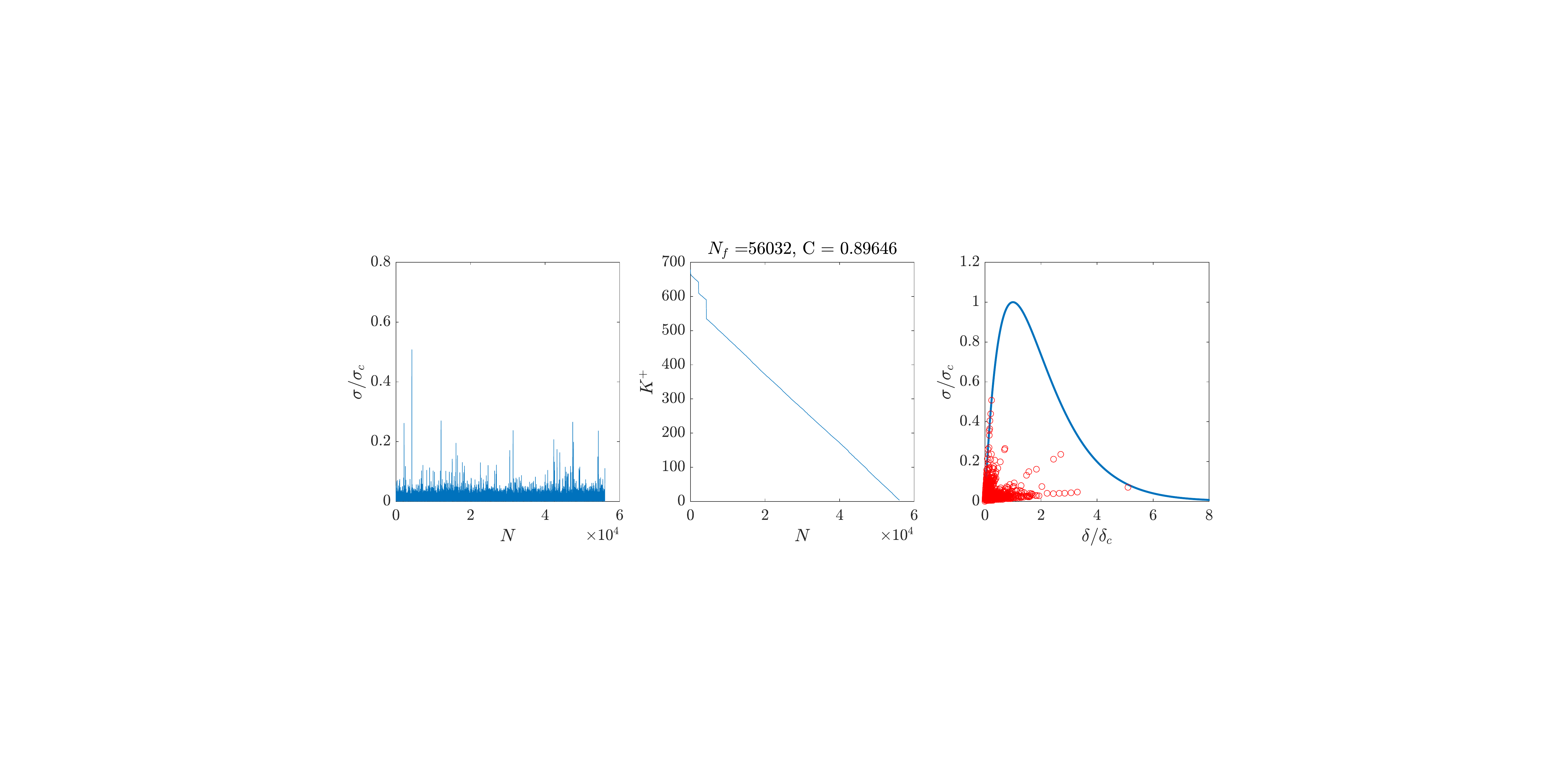} 
\includegraphics[trim=270 265 270 235,width=0.99\linewidth]{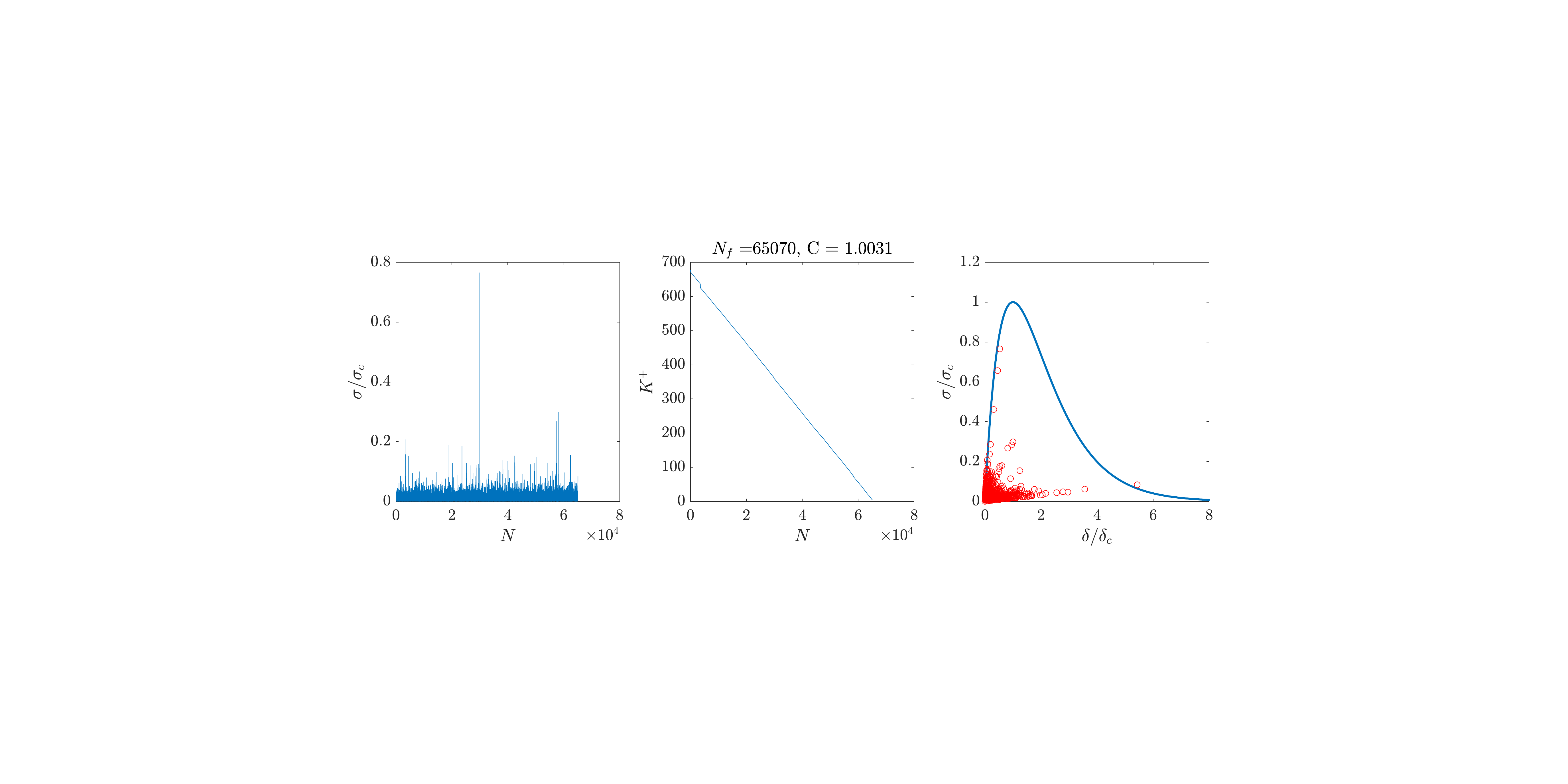} 
\end{center}
\caption{Two loading scenarios drawn from the same stochastic system and
for the same material parameters (left). The cohesive envelope (blue curves)
together with the peaks of the opening displacement, $\delta(t)$, are shown
(right).   Although, in the
first loading case the peaks are larger compared with the second case,
yet the material fails with a smaller number of cycles.
This early failure cannot be captured by the rainflow counting algorithm
which underestimates the fatigue damage. }
\label{fig4}
\end{figure}

In Figure \ref{fig4} we present two random samples (first column) of the load produced
by the system (\ref{driv1}). We can clearly observe the intermittent character
of the time series. On the right column we present the cohesive envelope (\ref{cohevn})
together with the local maxima of the opening displacement (red circles).
It is important to note that the two loading time series are qualitatively
similar. However, while in the first case the load has a much larger maximum
(grater than $0.75\sigma_{c}$) and several other intense peaks, the predicted
number of cycles is $N_{f}=65070$, close to the one predicted by the standard
rainflow counting prediction, since for this number of cycles we have $C=1.0031$, i.e. we have failure based on the Palmgren–Miner rule.
On the other hand, for the second
realization of the load, the peaks of the extreme events are much smaller
in magnitude (the larger is close to $0.5\sigma_{c}$), yet we have fatigue-crack
nucleation much earlier, $N_{f}=56032$. We emphasize that this material
failure
is inherently connected with the time-history of the load and cannot be captured
by the rainflow-counting algorithm which predicts a significantly larger
number of
cycles until material failure. Specifically, for this number of cycles the Palmgren–Miner rule gives $C=0.89<1$.



\section{Analytical approximation of the failure time pmf}
\label{sec:analytic-pdf}


If not for intersections between the $(\delta(t), \sigma(t))$ curve and the coherent envelope, the functional $N_f(\sigma)$ would be approximated the linear relationship (see eq. (\ref{eq_noup}))

\begin{equation}
\label{eq:n-formula-linear}
        N_f = \frac{K^+_0}{\Delta K}
\end{equation}

\noindent where $K^+_0$ is the initial stiffness and $\Delta K$ is the expected change in $K^+$ over a typical cycle corresponding to the load distribution. However, the coherent envelope changes things, by adding three effects:

\begin{enumerate}
\item Extremely large loads $\sigma_n > \sigma_c$ that immediately cause material failure.\item Up-crossings with the ascending part of the envelope, which cause jumps (discontinuities) on the evolution of $K^+$.
\item Up-crossings with the descending part of the envelope, which cause failure before $K^+$ has reached $0.$
\end{enumerate}
Each of these effects will be considered in order to derive an analytical approximation for the pmf of the failure time.
This is essential as a probabilistic quantification based on Monte-Carlo methods  require, in order to accurately resolve a pmf $f$ that take takes values as small as $\alpha$, order of $O(\frac{1}{\alpha})$ samples.  While there are techniques to slightly improve on these numbers (e.g., importance sampling), this is a critical barrier of Monte-Carlo ideas to estimate the long tail for rare failure events.

\subsection{Setup}

We will write the coherent envelope, introduced in Section \ref{sec:analytical-so-model}, in the generic form
\begin{equation}
        \sigma = \mathcal F(\delta),
\end{equation}

\noindent which has both a monotonic ascending and monotonic descending branch:
\begin{align*}
        \frac{\partial  \mathcal F}{\partial \delta} & > 0, \qquad \delta < \delta_c \\
        \frac{\partial \mathcal F}{\partial \delta} & < 0, \qquad \delta > \delta_c 
\end{align*}

\noindent and two corresponding inverses:  $ \mathcal F^{-1}_{asc}(\sigma)$ and $ \mathcal F^{-1}_{des}(\sigma)$.
Additionally, for what follows we define the functions:
\begin{equation}
        \kappa(\sigma) \triangleq \frac{\sigma}{ \mathcal F_{asc}^{-1}(\sigma)} \ \ \text{and} \ \ \ \eta(\sigma) \triangleq \frac{\sigma}{ \mathcal F_{des}^{-1}(\sigma)}
\end{equation}

\noindent which are assumed to be monotonic. These functions express the stiffness induced by an up-crossing with the ascending/descending part of the envelope and will be essential for our analysis. The monotonicity requirement is satisfied by typical coherent envelopes (e.g. eq. (\ref{cohevn})).

\subsection{Load statistics}
\label{sec:pdf-modeling-load}

We will assume that in the absence of envelope up-crossings the material stiffness, $K_n^+$, evolves linearly with the number of cycles and with its gradient given by the mean value (see eq. (\ref{eq:delta-k-approx})), which has been estimated through one of the methods detailed in Section \ref{slope}. This is a valid assumption as long as the spread of values of $\Delta \sigma_n^+$ is not systematically large. If this is not the case one can adopt a more complex model for the case of no envelope up-crossing using e.g. the Palmgren–Miner rule. 

Further, we assume known pdf of local load maxima, $\sigma_n^+$, as well as a known cumulative distribution:
\begin{align*}
        F_{\sigma^+}(\sigma) & = \int_{-\infty}^\sigma f_{\sigma^+}(s) ds.
\end{align*}
Finally, we will assume the \textbf{independent spike hypothesis}: that the amplitude of local maxima are uncorrelated.  This is \textit{not} true in general for narrow-banded processes, but it \textit{is} approximately true for `large enough' maxima, which is what is needed for our analysis.  This will be a key assumption in determining the probability distribution for several intermediate quantities below.  
\subsection{Failure time distribution}

\subsubsection{Damage due to terminal loads larger than $\sigma_c$}

The maximum value of $\mathcal F(\delta)$ is given by $\sigma_c$, and is the maximum load the material can sustain.  When the load on the material exceeds $\sigma_c$, it will fail no matter the fatigue history. We call these loads \textit{terminal}.
The failure cycle for this terminal mechanism is given by
\begin{equation}
        n_{x} = \argmin_{i > 0}\{\sigma^+_i > \sigma_c\}
\end{equation}
This is the expression for the first crossing time for the threshold $\sigma_c$.  Following the independent spike hypothesis, the probability of seeing an extreme load above a given threshold may be modeled by sequential Bernoulli trials.  In this case the pmf  of the first cycle $n_{x}$ when we have a spike of critical magnitude follows a geometric distribution
\begin{align}
        p_{N_x}(n) & =(1-p_{c})^{n-1}p_c,  \ \ \ n =1,2,...
\label{eq:extreme-pdf}
\end{align}
where,
\begin{align}        
        p_{c}  = P[\sigma_i > \sigma_c]   = \int_{\sigma_c}^{\infty} f_{\sigma^+}(s) ds.
\end{align}

\subsubsection{Damage due to up-crossings of the ascending part of the envelope}

\begin{figure}[b]
\centering
\includegraphics[width=0.92\linewidth]{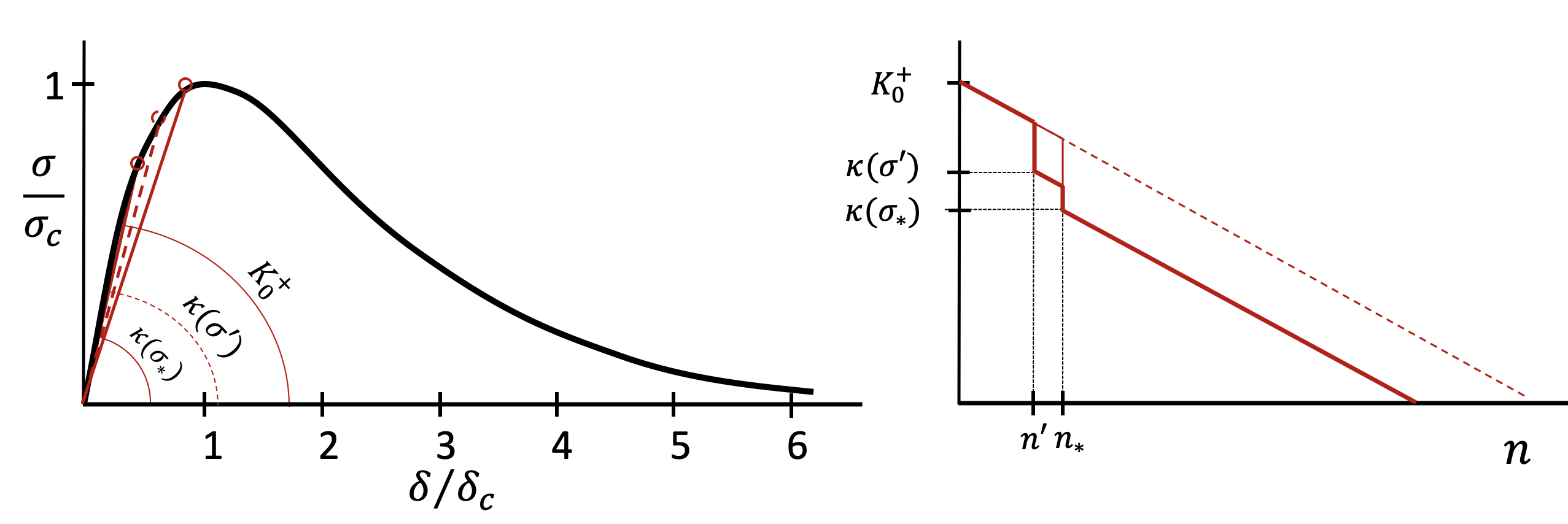}
\caption{Two loading scenarios shown in terms of the coherent envelope up-crossings and the stiffness in terms of the number of loading cycles.}
\label{fig_diag}
\end{figure}

In general, the graph of $(\delta_n, \sigma_n)$ may have multiple intersections
with the ascending part, leading to multiple discontinuous jumps in $K^+_n$
(Figure \ref{fig_diag}).  However, as we show below, the total fatigue lifetime effect depends only on the \textit{last} such intersection with the ascending part of the envelope and at what cycle this occurs.

\textbf{Invariance of the damage with respect to intermediate up-crossings}.
To prove this property we consider two scenarios where we have the same initial $K^{+}_0$. The last point where we have intersection of the ascending part of the coherent envelope is assumed to be the same for both scenarios (having slope $\kappa(\sigma^{+}_*)$), occurring at the same cycle, $n=n_{*}$. In the first scenario we have an additional intersection of the envelope, at slope $\kappa({\sigma^+}')$, occurring at some cycle, $n=n'<n_*$, while in the second case the only intersection occurs at $n=n_{*}$ (Figure \ref{fig_diag}). As we observe, for the first loading scenario, we have a jump at $n=n'$. Using simple geometry, the number of lost cycles due to this jump is given by (see Figure \ref{fig_diag} right)
\begin{equation}
\phi_{a1} \triangleq\ \frac{K^+_0 -n'\Delta K- \kappa({\sigma^+}')}{\Delta K}. 
\end{equation}
Using the same argument we also obtain the number of lost cycles due
to the second jump occurring at $n=n_{*}$: 
\begin{equation}
\phi_{a2} \triangleq\ \frac{\kappa({\sigma^+}')-(n_{*}-n')\Delta K- \kappa({\sigma^+}_*)}{\Delta K}. 
\end{equation}
Finally, by direct computational of the total number of cycles, we observe that this is equal to the lost cycles in the second loading scenario: 
\begin{equation}
\phi_{b} \triangleq\  \frac{K^+_0 -n_*\Delta K- \kappa(\sigma^+_*)}{\Delta K}=\phi_{a1}+\phi_{a2}. 
\end{equation}
The above argument can be generalized for an arbitrary number of intermediate jumps and proves that the damage due to intersections with the ascending part of the envelope depends only on the last such intersection. 
\textbf{\\ Damage quantification due to up-crossings of the ascending part for random loading}.
We have concluded that the number of lost cycles due to up-crossing of the ascending part of the envelope does not depend on intermediate up-crossings but only on the last up-crossing of the ascending curve.
To quantify this damage we define the \textit{damage quotient} $\phi(\sigma,n)$,
\begin{equation}
\phi(\sigma,n)\triangleq\frac{K^+_0 - \kappa(\sigma)}{\Delta K}-n.
\label{dam_eq}
\end{equation}
which is meaningful only when it is positive, i.e. only when we have an up-crossing of the envelope. \textit{The damage quotient essentially measures the magnitude of each jump on the material stiffness, expressed in a number of cycles lost due to this jump, every time we have an up-crossing}. For a generic loading sequence with peaks $\sigma_n^+, n=1,2, ...$ the number of lost cycles due to up-crossing with the ascending part of the envelope will be given by the maximum of this quantity:  
\begin{equation}
        n_{a}\left(\left\{\sigma^+_i \right\}_{i=1}^{\infty}\right) = \argmax_{0 \leq n < \infty} \phi(\sigma^+_n,n)= \argmax_{0 \leq n < \infty}  \left( \frac{K^+_0 - \kappa(\sigma^+_{n})}{\Delta
K}-n \right),
\end{equation}
with the condition that this maximum is a positive number, i.e. we have at
least one up-crossing with the ascending part of the envelope. If no up-crossing occurs then $n_a=0$.

The constant $K_0^+$ used in the normalization of $\phi(\sigma, n)$ is the `initial stiffness,' and corresponds to an infinitesimal first load peak.  Its value is given by
(eq. (\ref{init_s})),\begin{equation}
        K_0^+ = \lim_{\sigma \rightarrow 0} \kappa(\sigma),
\end{equation}

\noindent that is, the slope of the coherent envelope near the origin.

To quantify the pmf for $n_a$, we will begin by using the pdf  of the load peaks, $f_{\sigma^+}(a)$, and the monotonicty of function $\kappa(\sigma)$ to obtain the pdf for $\kappa(\sigma^+)$:
\begin{equation}
f_{\kappa}(a)=\frac{f_{\sigma^+}(\kappa^{-1}(a))}{|\kappa'(\kappa^{-1}(a))|}.
\end{equation}
Based on this pdf we now obtain the pdf for the damage quotient $\phi(\sigma^+_n,n)$:
\begin{equation}
f_{\phi_n}(a)=\Delta Kf_{\kappa}(K_0^+-n\Delta K-a\Delta K).
\end{equation}
and the corresponding cdf:
 \begin{equation}
F_{\phi_n}(a)=1-F_{\kappa}(K_0^+-n\Delta K-a\Delta K).
\end{equation}
For any sequence of independent random variables $x_1, x_2, ..., x_N$ with corresponding pdf $f_1, f_2, ..., f_N$, and cdf $F_1, F_2, ..., F_N$, we can show, using derived distributions, that the cumulative distribution function of the maximum is given by 
\begin{equation}
F_{\max}(x)=\prod_{\substack{j=1 }}^N F_j(x),
\end{equation}
while using a total probability argument we can prove that the pmf for the argument of the maximum is given by \cite{Habibi}
\begin{equation}
p_{\max}(n)=\int_{-\infty}^{\infty} \prod_{\substack{j=1 \\ j\neq n}}^N F_j(x) f_n(x)dx=\int_{-\infty}^{\infty} F_{\max}(x)
\frac{f_n(x)}{F_n(x)}dx, \ \ n=1,...,N
\end{equation}
Applying this result for our problem leads to,
\begin{equation}
p_{n_a}(n)=\int_{-\infty}^{\infty} Q_{\kappa}(x\Delta K) \frac{
\Delta Kf_{\kappa}(K_0^+-n\Delta K-x\Delta K)}{1-F_{\kappa}(K_0^+-n\Delta K-x\Delta K)}dx, \ \ n=1,...,
\label{pmf1}
\end{equation}
where $Q_{\kappa}$ is a function that is independent of $n$ (i.e. it has to be computed once):
\begin{equation}
Q_{\kappa}(y)=\lim_{n\rightarrow\infty}\prod_{\substack{j=1}}^n \left(1-F_{\kappa}(K_0^+-j\Delta K-y)\right).
\end{equation}
Note that
\begin{equation}
F_{\phi_{n_a}}(x)=Q_{\kappa}(x\Delta K).
\label{eq:Q}
\end{equation}
To compute the function $Q_{\kappa}(y)$ we consider its logarithm
\begin{equation}
\log(Q_{\kappa}(y))=\sum_{j=1}^{\infty}\log \left(1-F_{\kappa}(K_0^+-j\Delta
K-y)\right).
\end{equation}
We multiply both sides of the equation with $\Delta K$ and consider the limit of $\Delta K \rightarrow 0$ as for typical applications $\Delta K$ has a very small value which corresponds to a large number of cycles until failure. This allows us to express the right hand side in terms of an integral.
\begin{equation}
\lim_{\Delta K\rightarrow0}\left[\log(Q_{\kappa}(y))\Delta K\right]=\int_0^{\infty}\log \left(1-F_{\kappa}(K_0^+-y-z)\right)dz.
\end{equation}
Therefore, for finite (and small) $\Delta K$ we have
\begin{align}
\begin{split}
Q_{\kappa}(y)& =\exp\left(\frac{1}{\Delta K}\int_0^{\infty}\log
\left(1-F_{\kappa}(K_0^+-y-z)\right)dz\right) \\ 
& =\exp\left(\frac{1}{\Delta K}\int_y^{\infty}\log
\left(1-F_{\kappa}(K_0^+-z)\right)dz\right).
\end{split}
\label{eq:q1}
\end{align}
Substituting the above into (\ref{pmf1}) results in the pmf of the cycles until the last up-crossing of the ascending part of the envelope:
\begin{equation}
p_{n_a}(n_{a})=\int_{-\infty}^{\infty} Q_{\kappa}(x\Delta K)W_{\kappa}( n_{a}\Delta K+x\Delta K) \Delta
Kdx, \ \ n_{a}=1,...,
\label{pmf2}
\end{equation}
where,
\begin{align}
W_{\kappa}( y)&\triangleq\frac{
f_{\kappa}(K_0^+-y)}{1-F_{\kappa}(K_0^+-y)}.
\label{eq:W}
\end{align}

\subsubsection{Damage due to up-crossings of the descending part of the envelope}

Any intersection with the descending part of the coherent envelope will immediately cause material failure.  As such, there can only be one such intersection.  In order to quantify the statistics of this event we define the \textit{anticipation function} $\psi$:

\begin{align}
\begin{split}
        \psi(\sigma_{n}, n)  & \triangleq\eta(\sigma_n)- K_n^+ , \\
        K_n^+ & = K^+_{n_a} - (n - n_a) \Delta K
\end{split}
\end{align}
where $\eta(\sigma)=  \frac{\sigma}{ \mathcal F^{-1}_{des}(\sigma)}$, $K^+_n$ is the material stiffness coefficient before cycle $n$, and 
$K^+_{n_a}=\kappa(\sigma_{n_a})$. The material stiffness can be expressed in terms of the damage quotient (eq. (\ref{dam_eq})) and its maximum value as follows:
\begin{equation}
K_n^+ = K_0^+-(\phi_{n_a}+n)\Delta K,
\end{equation}

where the pdf for $\phi_{n_a}$ is given by eq. (\ref{eq:Q}). The material failure time is given by the first zero up-crossing of the anticipation function: 
\begin{equation}
        N_{f}= \min \{n:  \psi(\sigma_{n}, n)=\eta(\sigma_n)- K_0^++(\phi_{n_a}+n)\Delta K> 0 \}.
\end{equation}

We first compute the pdf for $\eta$. This will be given by:

\begin{equation}
f_{\eta}(a)=\frac{f_{\sigma^+}(\eta^{-1}(a))}{|\eta'(\eta^{-1}(a))|}.
\end{equation}Therefore, conditioning on $\xi=\phi_{n_a}\Delta K$, which expresses the maximum lost stiffness of the material due to an up-crossing with the ascending part of the envelope, and $n_a$,  the cycle when this up-crossing occurs, we will have

\begin{equation}
F_{\psi_n}(a|\xi,n_{a})= 1 - F_{\eta}(a+K_0^+-\xi-n_a\Delta
K).
\end{equation}To this end, the probability of having a material failure at $N_{f}$ cycles is
\begin{align}
\begin{split}
p_{N_{f}}(n|\xi,n_{a})&= \left(1 - F_{\eta}(K_0^+-\xi-n\Delta
K)\right) \prod_{m=n_{a}+1}^{n-1}F_{\eta}(K_0^+-\xi-m\Delta
K), \\ n&=n_{a}
+1,n_{a}+2,...
\end{split}
\end{align}
where, $\xi$ follows the cdf $Q(\xi)$ (eq. (\ref{eq:Q}) and (\ref{eq:q1})), while $n_a$ follows the pmf in eq. (\ref{pmf1}). We consider the logarithm of this equation:
\begin{align*}
\begin{split}
\log p_{N_{f}}(n|\xi,n_{a})&=\log \left(1 - F_{\eta}(K_0^+-\xi-n\Delta
K)\right) \\ &+\sum_{m=n_{a}+1}^{n-1}\log\left(F_{\eta}(K_0^+-\xi-m\Delta
K) \right).
\end{split}
\end{align*}
For the last term we set
\begin{equation}
\log V_{\eta}\triangleq\ \sum_{m=n_{a}+1}^{n-1}\log\left(F_{\eta}(K_0^+-\xi-m\Delta
K) \right).
\end{equation}We multiply and divide the right hand side with $\Delta K$ and express it as an integral

\begin{align}
\log V_{\eta} & =\frac{1}{\Delta K}\int_{n_a\Delta K}^{n\Delta K}\log\left(F_{\eta}(K_0^+-\xi-s) \right)ds  =\frac{1}{\Delta K}\int_{n_a\Delta K+\xi}^{n\Delta K+\xi}\log\left(F_{\eta}(K_0^+-s)
\right)ds, \\
V_{\eta} (u, v) & = \exp \left( \frac{1}{\Delta K}\int_{u}^{v}\log\left(F_{\eta}(K_0^+-s)
\right)ds  \right),
\label{eq:V}
\end{align}
where the approximation is based on the assumption of small $\Delta K$ and large number of cycles until failure.  Going back to the pmf for $N_f$ we have 
\begin{align}
\begin{split}
p_{N_{f}}(n|\xi,n_{a})&=\left( 1 - F_{\eta}(K_0^+-\xi-n\Delta
K)\right) V_{\eta}(n_a\Delta K+\xi,n\Delta K+\xi), \\ n&=n_{a}
+1,n_{a}+2,...
\end{split}
\end{align}
The special case where no up-crossings with the ascending part of the envelope occur is when the magnitude of the maximum stiffness jump is zero, $\xi=0$ and $n_a=0$. In this case we will have\begin{equation}
p_{N_{f}}(n|\xi=0,n_{a}=0)=(1-F_{\eta}(K_0^+-n\Delta
K))V_{\eta}(0,n\Delta K), \ \ n=1,2,...
\end{equation} 
Using eq. (\ref{pmf1})
and assuming that probable $n$ is much larger than the probable values of $n_a$ (so we do not have to formally condition on $n>n_a$), as well as a small $\Delta K$, we have
\begin{align*}
\begin{split}
p_{N_{f}}(n|\xi)& =\sum_{n_a=1}^{\infty}p_{N_{f}}(n|\xi,n_{a})p(n_a) \\
& = \left( 1 - F_{\eta}(K_0^+-\xi-n\Delta K) \right).   \\
& \int_{-\infty}^{\infty}  \sum_{n_a=1}^{\infty}Q_{\kappa}(x\Delta K)V_{\eta}(n_a\Delta K+\xi,n\Delta K+\xi)W_{\kappa}(n_a\Delta K+x\Delta K)\Delta Kdx \\
& = \left( 1 - F_{\eta}(K_0^+-\xi-n\Delta K)\right) \int_{-\infty}^{\infty} 
 \int_{0}^{\infty} Q_{\kappa}(x\Delta K) V_{\eta}(\zeta+\xi,n\Delta K+\xi)W_{\kappa}(\zeta+x\Delta K)d\zeta dx\\
& =\frac{1}{\Delta K}\left( 1 - F_{\eta}(K_0^+-\xi-n\Delta K) \right) \int_{-\infty}^{\infty} 
 \int_{0}^{\infty} Q_{\kappa}(y)V_{\eta}(\zeta+\xi,n\Delta K+\xi)W_{\kappa}(\zeta+y)d\zeta dy.
\end{split}
\end{align*}
Finally, we integrate over the variable $\xi$ after we multiply with the corresponding pdf $Q_{\kappa}'(\xi)$:
\begin{align*}
p_{N_{f}}(n)=\frac{1}{\Delta
K}\int_{-\infty}^{\infty} \left( 1 - F_{\eta}(K_0^+-\xi-n\Delta K) \right) Q_{\kappa}'(\xi)\int_{-\infty}^{\infty} 
 \int_{0}^{\infty} Q_{\kappa}(y)V_{\eta}(\zeta+\xi,n\Delta K+\xi) W_{\kappa}(\zeta+y)d\zeta
dyd\xi.
\end{align*}
This expression can be also written in a more compact form
\begin{equation}
p_{N_{f}}(n)=\frac{1}{\Delta
K}\int_{-\infty}^{\infty} \int_{0}^{\infty} \left( 1 - F_{\eta}(K_0^+-\xi-n\Delta K) \right) V_{\eta}(\zeta+\xi,n\Delta K+\xi)Q_{\kappa}'(\xi)S_{\kappa}(\zeta) 
 d\zeta
d\xi.
\label{eq:final}
\end{equation}
where,

\begin{equation}
S_{\kappa}(\zeta)\triangleq\int_{-\infty}^{\infty}Q_{\kappa}(y)W_{\kappa}(\zeta+y)dy.
\label{eq:helper-s}
\end{equation}
Note that the integrands in equations (\ref{eq:final}) and (\ref{eq:helper-s}) have compact support, subsets of the interval $[0, K_0^+]$.
Expression (\ref{eq:final}) together with the functions $V_{\eta}$ (eq. (\ref{eq:V})), $Q_{\kappa}$ (eq. (\ref{eq:q1})), $W_{\kappa}$ (eq. (\ref{eq:W})) and $S_{\kappa}$ consist of a full approximation of the cycles until material failure. These functions are given in terms of the coherent envelope shape and load peak statistics.


\subsubsection{Combined failure time}


From equations (\ref{eq:extreme-pdf}) and (\ref{eq:final}), we have expressions for the distribution of $N_x$ and $N_f$, the failure times for the terminal-load case and quiescent-with-extremes-load pathways respectively.  To find the combined failure time, we simply check which comes first:
\begin{equation}
        N_{\mbox{tot}}(\sigma) = \min\{N_x, N_f\}
\end{equation}
The corresponding pmf is given by
\begin{equation}
        p_{N_{tot}}(n) = p_{N_f}(n) (1 - F_{N_x}(n)) + p_{N_x}(n) (1 - F_{N_f}(n))
\end{equation}
\noindent where $F_{N_x}(n)$ and $F_{N_f}(n)$ are the cumulative probabilities corresponding to $p_{N_x}(n)$ and $p_{N_f}(n)$ (from equations (\ref{eq:extreme-pdf}) and (\ref{eq:final}), respectively).  In general, the terminal-load pathway may only contribute to the far left of the distribution (extremely early failure times), while the quiescent-with-extremes-load pathway controls the rest of the shape.
However, in realistic applications it is very rare for a material to fail due to the terminal pathway as this mechanism can be easily accounted for in the design phase. 

\subsection{Comparison of analytic pmf with Monte Carlo simulations}

In order to compare the probabilistic framework developed in Section \ref{sec:analytic-pdf} to direct integration of the SO model we employ the coherent envelope shape in equation (\ref{cohevn}) with scale parameters $\sigma_c = \delta_c = 1$ and $\delta_a = 300$.  We consider realizations of a stochastic signal consisting of a zero-mean stochastic process superimposed with intermittent events randomly occurring and with amplitude given by a Rayleigh distribution. 

\begin{figure}[hbt]
\centering
\includegraphics[width=0.32\linewidth]{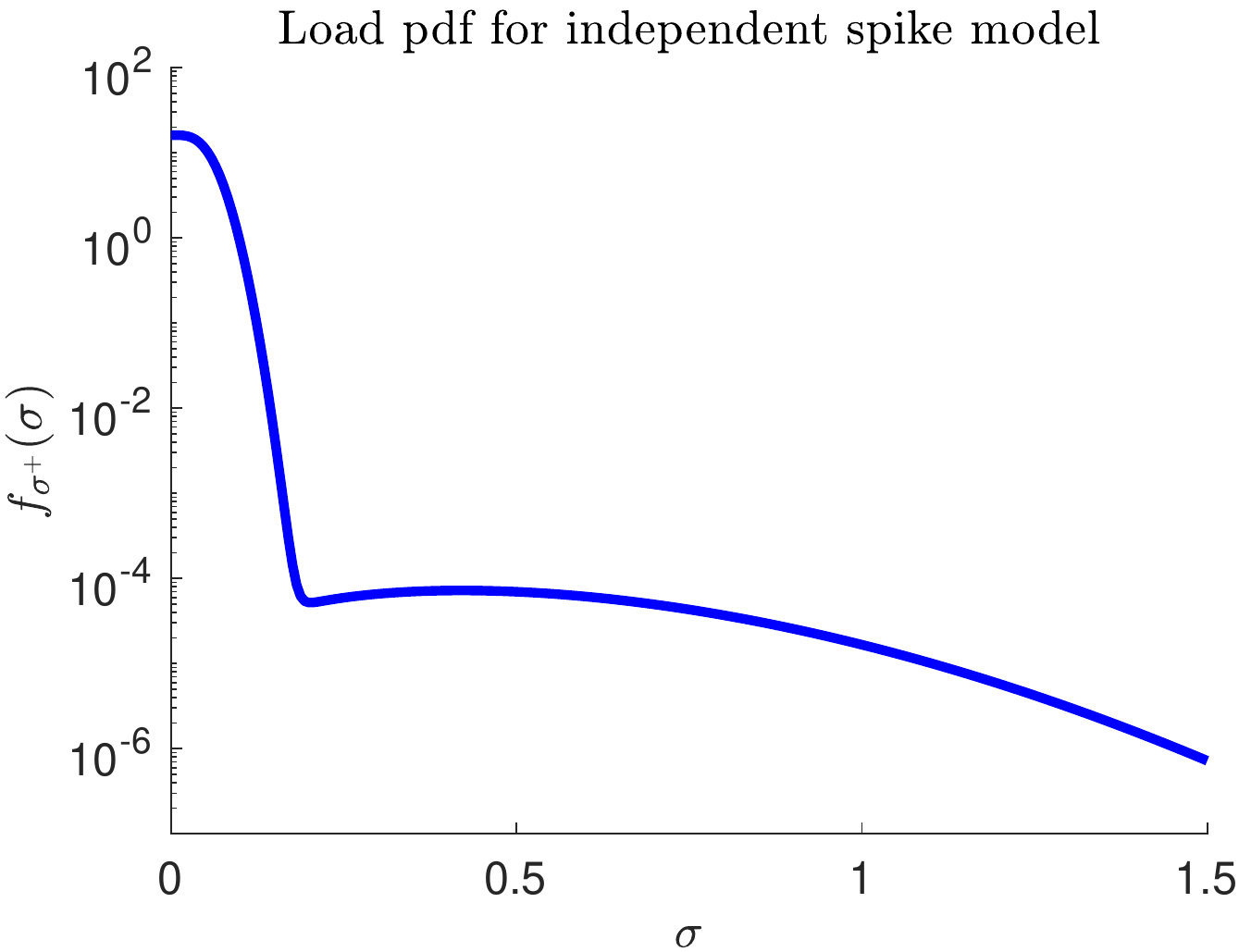}
\includegraphics[width=0.32\linewidth]{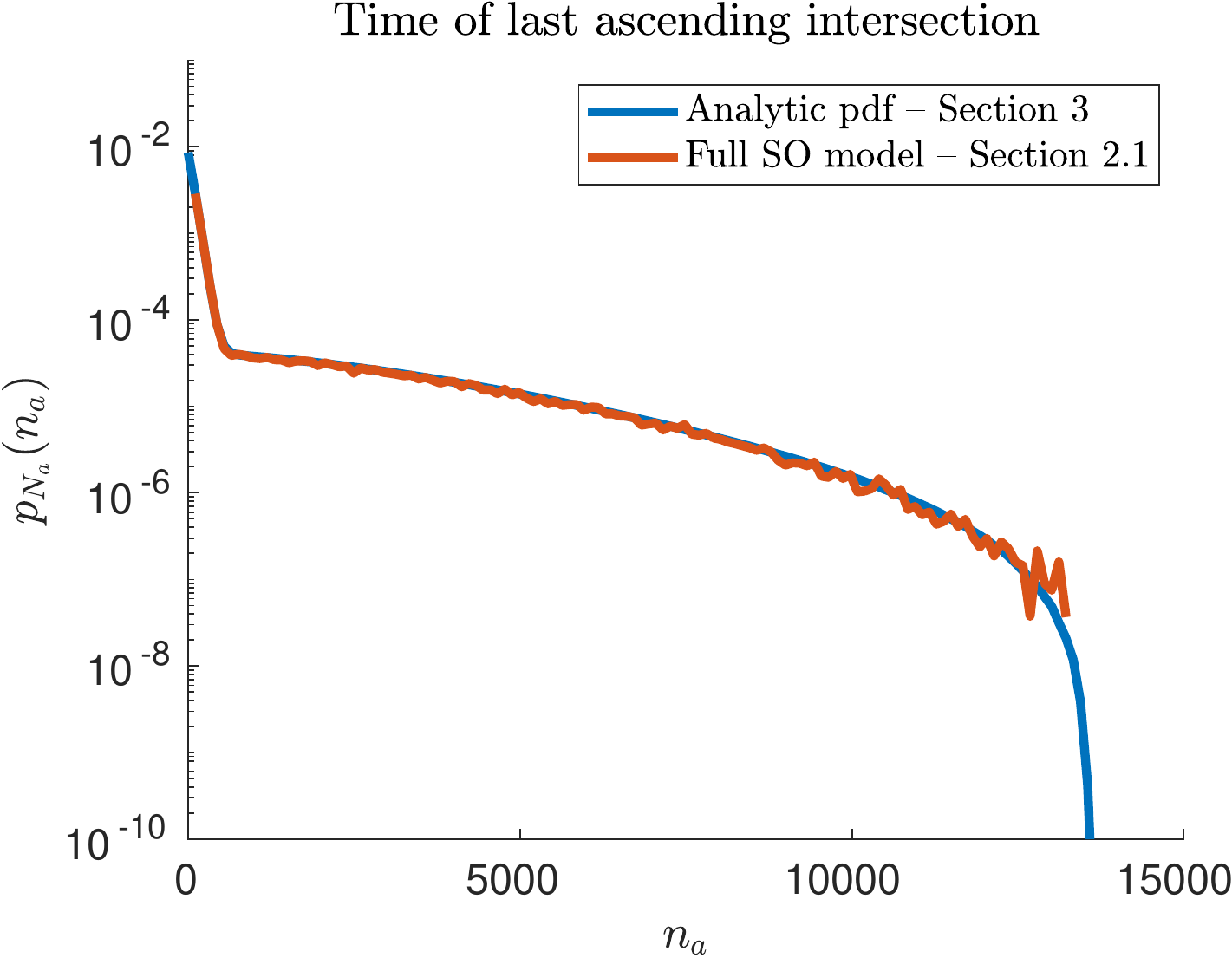}
\includegraphics[width=0.32\linewidth]{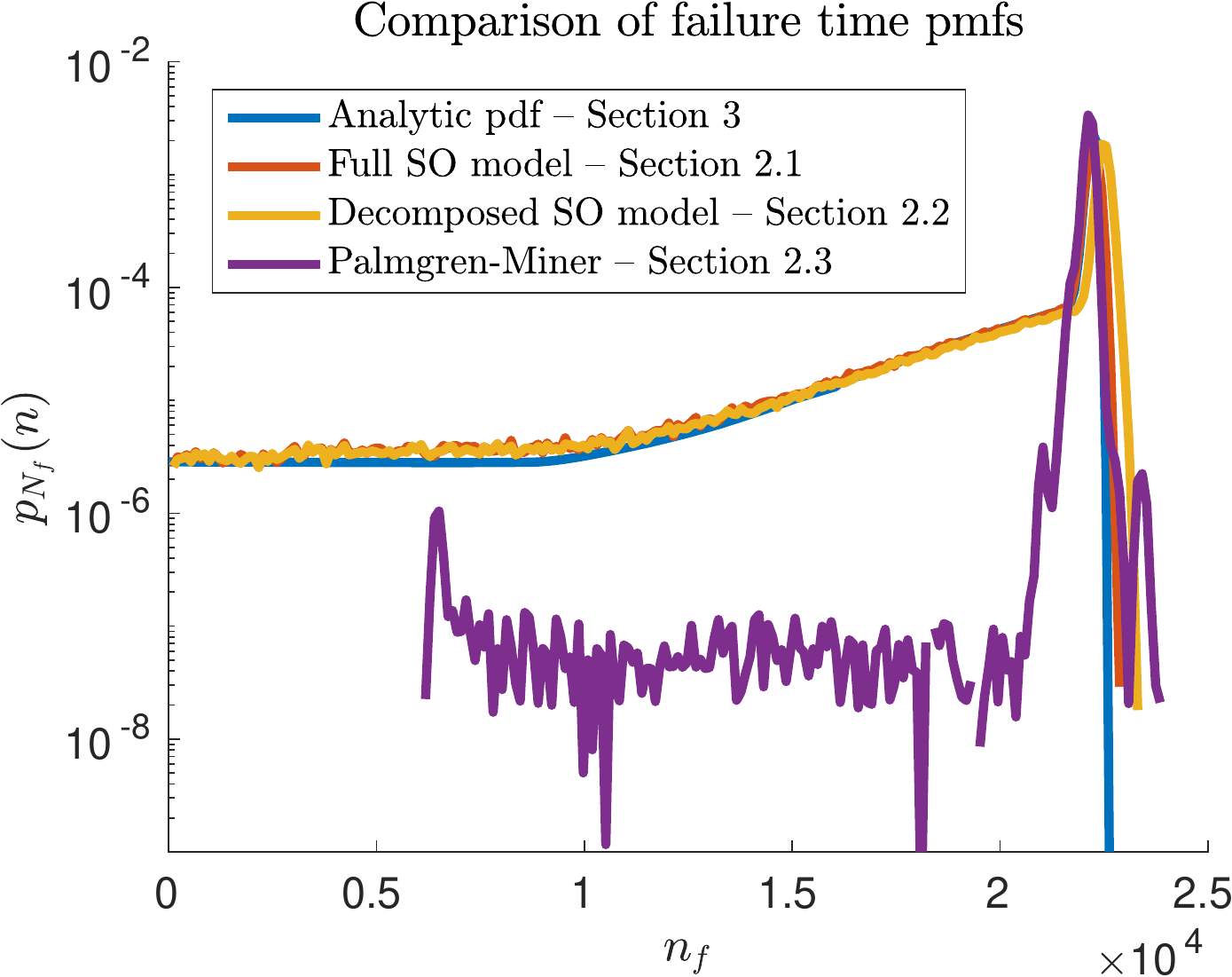}
\caption{Comparison of important distributions between the analytic framework
and Monte Carlo simulation for $p_b=1/20000$.  a)  The load peak ($\sigma^+$) pdf used for the independent spike model.  b) The pmf for the time of the last upcrossing, $N_a$, of the ascending branch of the coherent envelope.  c) The pmf for the time of fatigue-crack nucleation, $N_f$.}
\label{fig:analytic-mc-comparison}
\end{figure}

Specifically, for each peak we first run a Bernoulli trial with success probability $p_b$.  If this trial fails, then we have a non-extreme or \textit{quiescent} peak, in which case we draw a sample from an absolute normal distribution with mean, $\mu_q = 0.03$, and standard deviation, $\rho_q = 0.03$. Note that for any broad-band stochastic process the peaks follow a Gaussian distribution \cite{SooGrig}, but given that in the present context only the positive peaks matter, we consider an absolute normal distribution. If the Bernoulli trial succeeds, we instead draw an \textit{extreme} sample from a Rayleigh distribution with scale parameter, $r=0.42$.  This results a probability density function for the load-peak ($\sigma^+$) having a sharp peak near $\sigma^+ = 0.03$ (i.e., much less than $\sigma_c$), but a long right tail with a small but finite probability-mass corresponding to values of $\sigma$ larger than $\sigma_c$ (Figure \ref{fig:analytic-mc-comparison}a). 

In Figure \ref{fig:analytic-mc-comparison}b we present the comparison of the pmf for the cycle number associated with the last up-crossing of the ascending part of the envelope, $p_{N_a}$, while in Figure \ref{fig:analytic-mc-comparison}c we compare the failure time distribution between the analytic framework and Monte Carlo simulations, based both on the fast decomposition synthesis approximation and the full SO integration.  Additionally, the predictions of the Palmgren-Miner rule are shown for reference.  It is clear that the SO model underestimates the variance in failure time and cannot capture the heavy left tail.

\begin{figure}[h]
\centering
\includegraphics[width=0.42\linewidth]{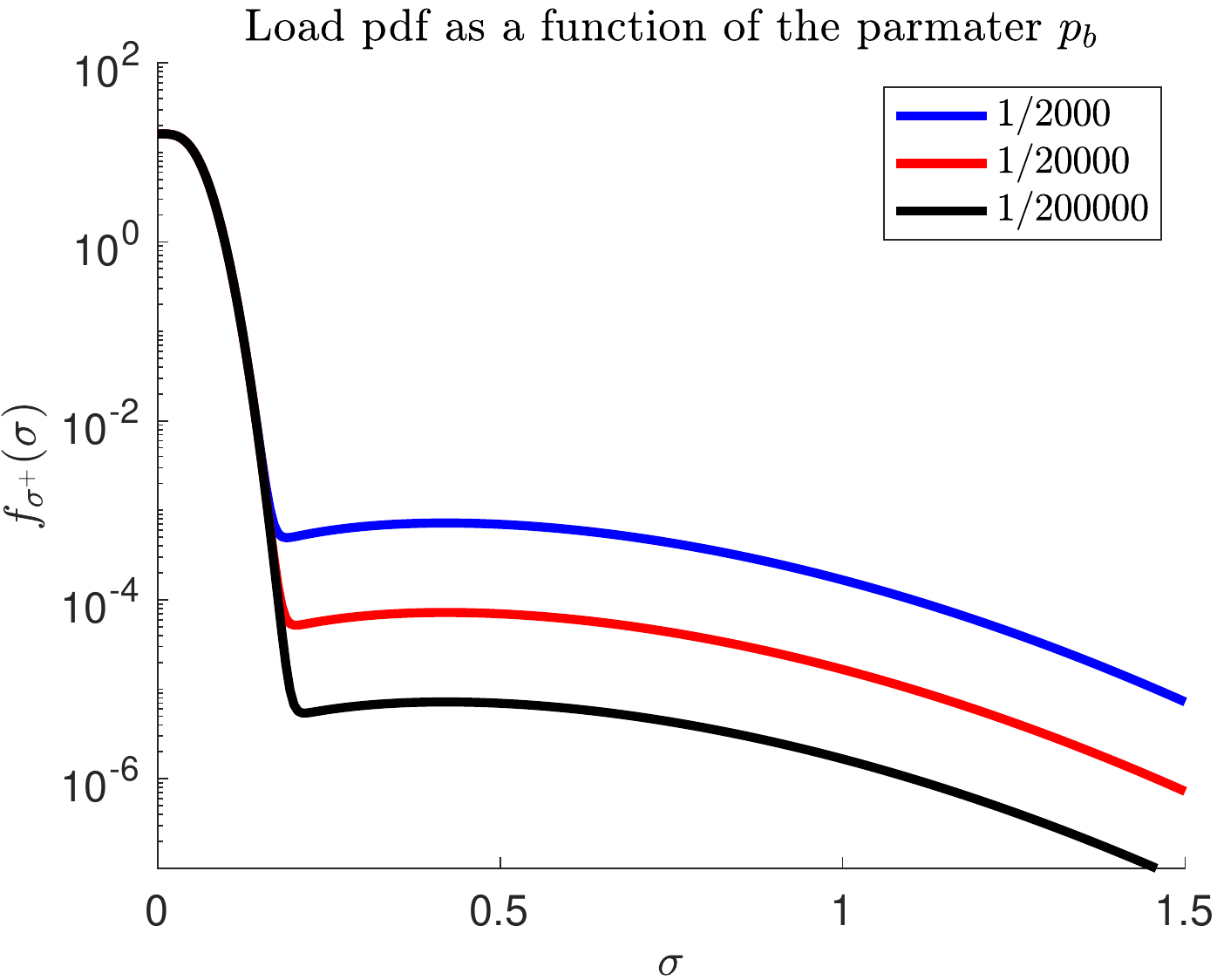}
\includegraphics[width=0.42\linewidth]{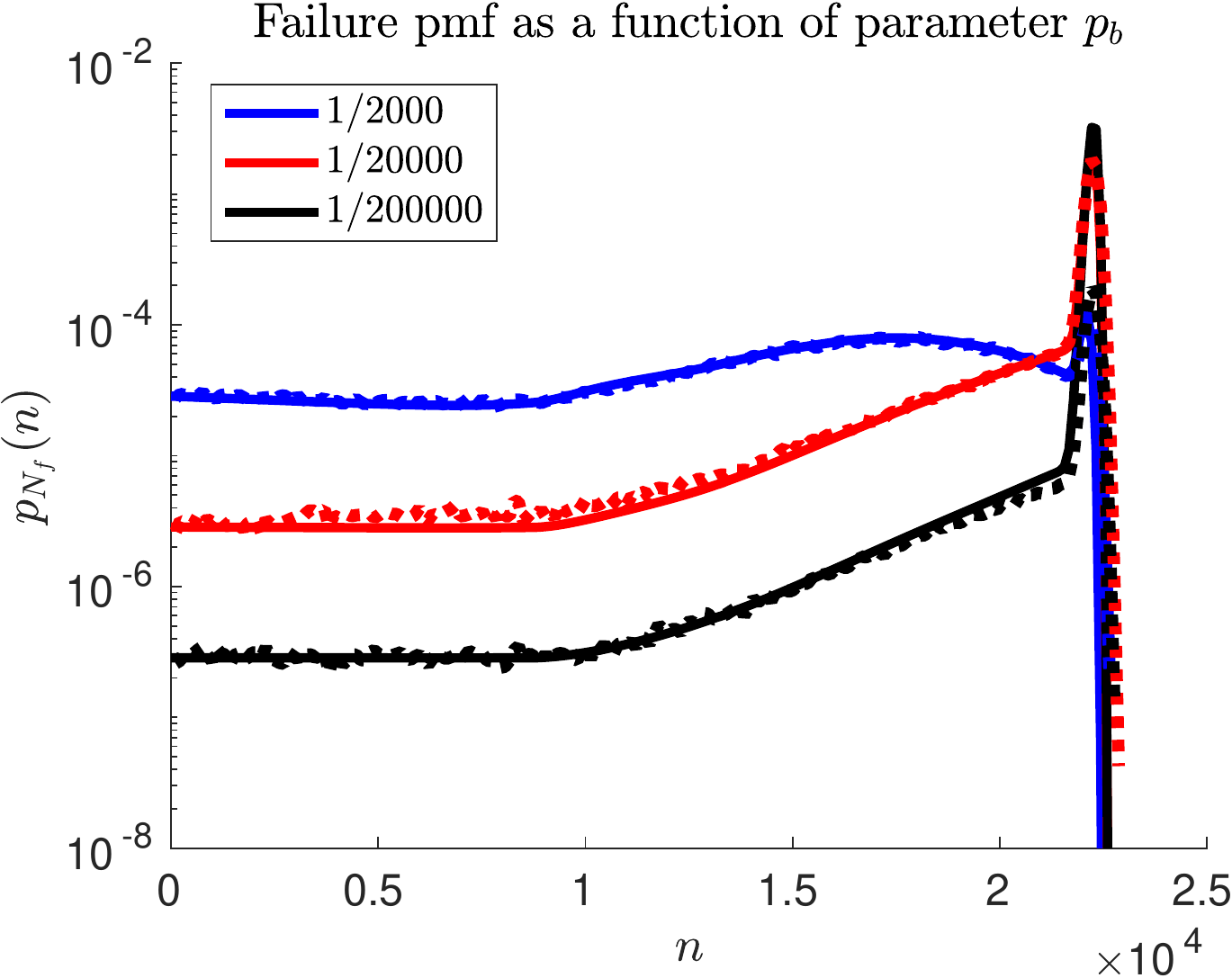}
\caption{Comparison of failure time pmf for different values of $p_b$ in the independent spike model  a) Load peak $(\sigma^+)$ probability density functions for different $p_b$.  b) Failure time distributions for different $p_b$. Analytical pmf is shown with solid curves and Monte-Carlo simulations with dashed curves.}
\label{fig:analytic-mc-comparison-p}
\end{figure}

In the preceding experiment, a given load signal may have zero, one, or more than one extreme peaks or Bernoulli successes (corresponding to a Rayleigh draw). The mean number of such peaks is given by $p_b \tilde{N}$, where $\tilde{N} \approx 2.3 \times 10^4$ is the mean fatigue lifetime.  To examine the effect of $p_b$ on the distribution of failure times, we repeat this experiment for three different values of $p_b$, given by $[1 / 2000,  ~ 1 / 20000, ~ 1 / 200000]$.  For the presented experiment, we use $1.2 \times 10^6$, $1.9 \times 10^6$, and $3.2 \times 10^6$ Monte Carlo realizations for increasingly small value of $p$. The corresponding load distributions, as well as failure time distributions, are displayed in Figure \ref{fig:analytic-mc-comparison-p}. We emphasize the favorable agreement between the analytical approximation of the pmf and the direct Monte Carlo simulations. It is clear that as extreme events become less likely we have less mass in the long left tail and the distribution becomes closer to the one predicted by the Palmgren-Miner rule.  However, as the extreme event rate rises, the long left tail begins to absorb probability mass from the sharp distribution mode.  

\section{Robustness with respect to the coherent envelope properties}

The preceding computations have been based upon the assumption of a known coherent envelope. However, in practice it may be hard to obtain the exact shape of the coherent envelope. In this section, we show that the precise functional form of the coherent envelope has only a small impact on the shape of the failure distribution, compared to the more readily measured quantities $\sigma_c$ and $\delta_a$ and the $SN$ curve.

To show this, we first emphasize certain geometric constraints on the two branches of the coherent envelope due to the monotonicity requirements on $\kappa(\sigma)$ and $\eta(\sigma)$.  These require that the ascending branch be monotonic increasing, and while they do not entirely rule out the possibility of an inflection point, there is a bound on the degree of non-concavity. Second, the material $SN$ curve imposes a constraint on the combination of the ascending and descending branches.  In particular, the $SN$ curve for a harmonic load in the SO model can be derived from the following relation  \cite{serebrinsky05, arias06},
\begin{equation}
\label{eq:inverse-gap}
        N = \frac{\delta_a}{1 - R} \left ( \frac{1}{\delta^-} - \frac{1}{\delta^+} \right)
\end{equation}

\noindent where, $R = \frac{\sigma_{min}}{\sigma_{max}}$, and $\delta^{\pm}$ is the solution to the descending and ascending branch of the coherent envelop for $\sigma=\sigma_{max}$.  The derivation of this relation assumes that the difference between successive $\delta$ is small (i.e., that $N$ is large), which is true when $\sigma < \sigma_c$.  In this case, the relationship may be used to derive the descending branch from a given form for the ascending branch and $SN$ curve, as given below:
\begin{equation}
\label{eq:inverse-gap-direct}
         \mathcal F^{-1}_{des}(\sigma) = \frac{1}{[ \mathcal F^{-1}_{asc}(\sigma)]^{-1} - \frac{N(\sigma)}{\delta_a}}
\end{equation}
This reduces the problem of fitting the entire coherent envelope to one of fitting only the ascending branch.  Considering the inverse ascending branch, $\mathcal F^{-1}_{asc}(\sigma)$, we require
the following properties

\begin{figure}[t]
\centering
\includegraphics[width=0.45\linewidth]{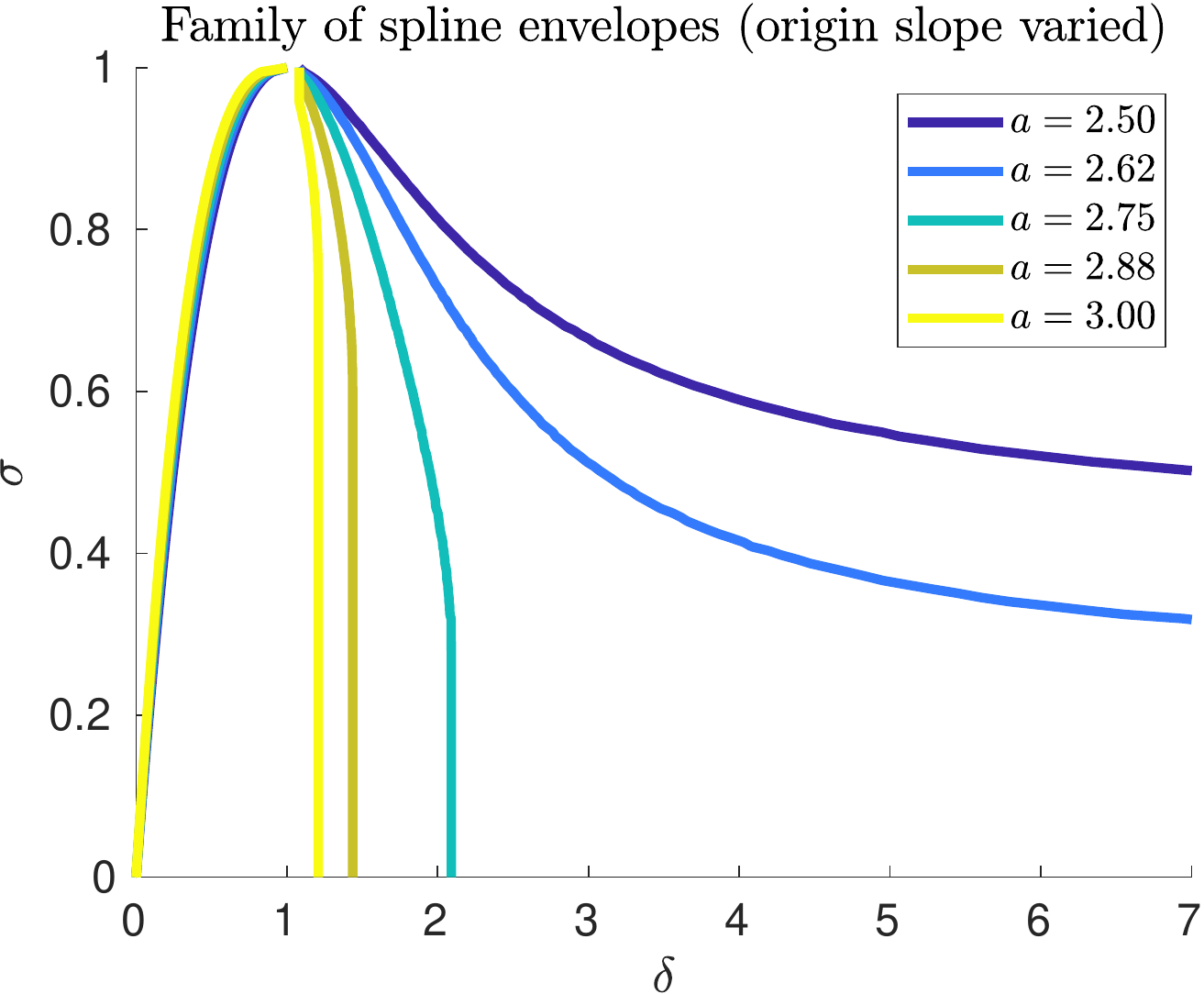}
\includegraphics[width=0.45\linewidth]{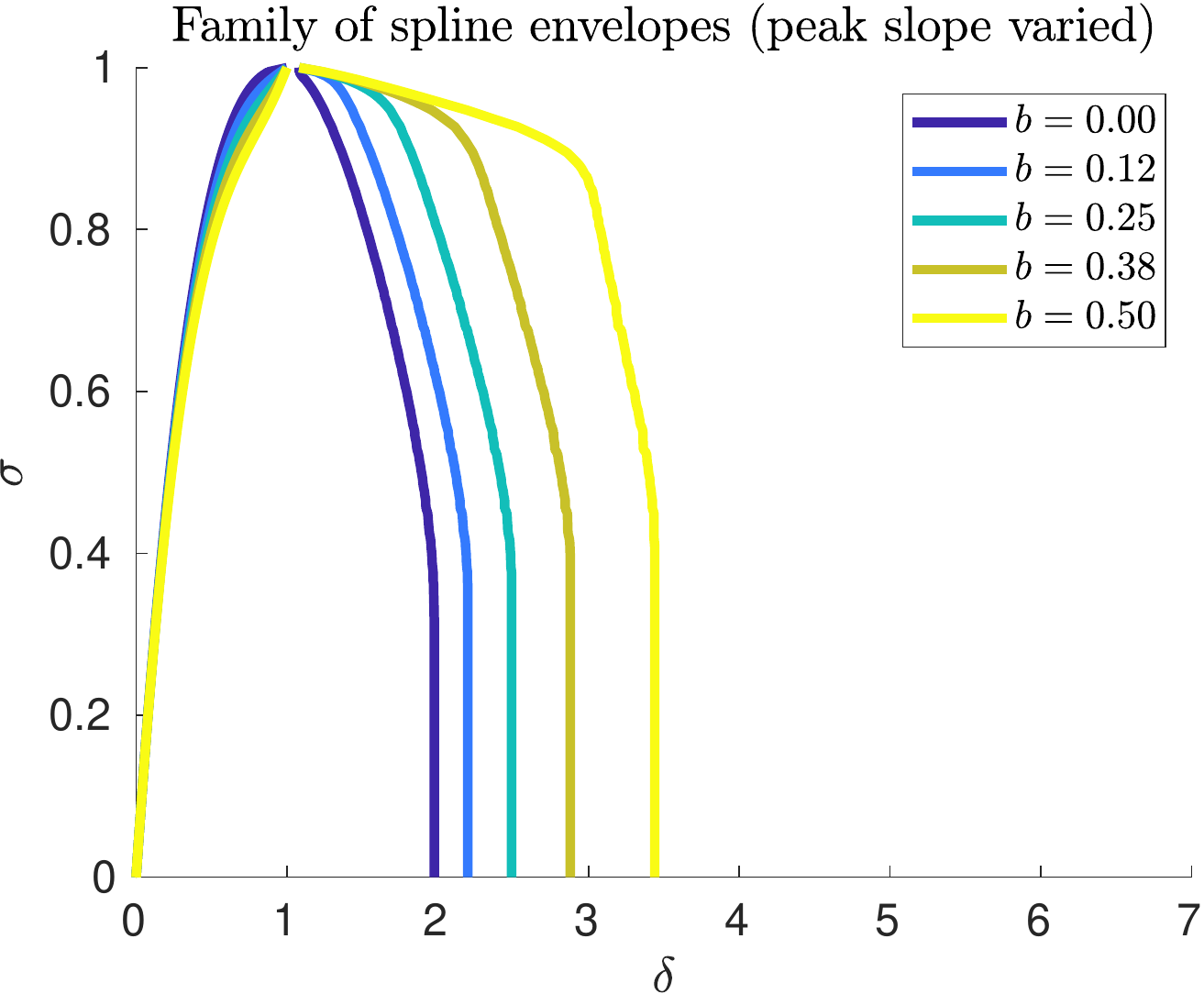}
\caption{Sample spline coherent envelopes, based on the same $SN$ curve as
the coherent envelope (\ref{cohevn}) (c.f. figure \ref{fig:so-sn-plot}).  a)  Varied slope at
$(0,0)$, corresponding to different values of $K_0^+$.  b)  Varied slope
at $(1,1)$.  The coherent envelope (\ref{cohevn}) is closely approximated
by the spline envelope with $(a, b) = (2.73, 0)$.}
\label{fig:spline-envelope-comparison}
\end{figure}

\begin{enumerate}
\item $ \mathcal F^{-1}_{asc}(0) = 0$,
\item $ \mathcal F^{-1}_{asc}(\sigma_c) = \delta_c$.
\end{enumerate}
These requirements naturally suggest a cubic spline parametrization, with two adjustable parameters, ($a,b$), corresponding to the slope at $(0, 0)$ and the slope at $(\delta_c, \sigma_c)$, respectively.   The slope at the origin, $a$, represents the value, $K_0^+$ (eq. (\ref{init_s})). The slope at the peak, $b$, allows for the    possibility of a discontinuous derivative between the two branches. Figure \ref{fig:spline-envelope-comparison} shows two families of splines that correspond to varied parameters, $a$ and $b$.

We emphasize that for some combinations of the ascending branch parameters and $SN$ relations, equation (\ref{eq:inverse-gap-direct}) allows for two solutions: either a horizontal asymptote in the descending branch, or a sweep back to the origin. The second case is not monotonically decreasing and therefore  we accept only the first solution where $ \mathcal F_{des}(\sigma)$ reaches a horizontal asymptote (Figure \ref{fig:spline-envelope-comparison}a). 

Two other important parameters are the scaling parameters $\sigma_c$ and $\delta_a$.  The critical stress $\sigma_c$ sets the scale between the typical load magnitude and the material maximum sustainable load (roughly, the onset of plastic deformation).  The endurance length $\delta_a$ (or more accurately, the ratio between the critical length $\delta_c$, assumed to be $1$, and the endurance length $\delta_a$) sets the scale of cycle fatigue, expressed as the rate of change of $K^+$; recall that in equation (\ref{eq:delta-k-approx}), the characteristic fatigue increment per cycle $\Delta K$ is inversely proportional to $\delta_a$.

\begin{figure}[h]
\centering
\includegraphics[width=0.45\linewidth]{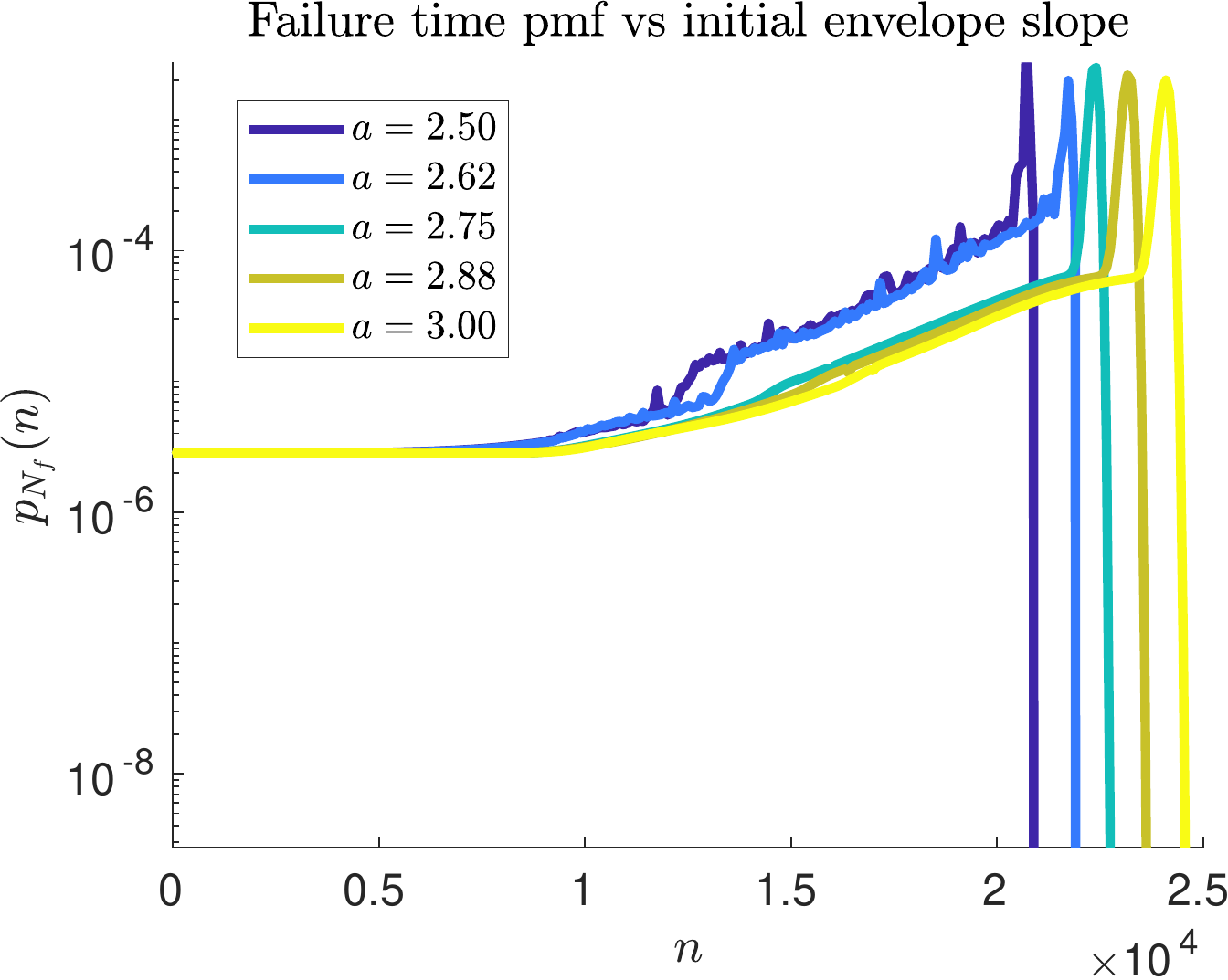}
\includegraphics[width=0.45\linewidth]{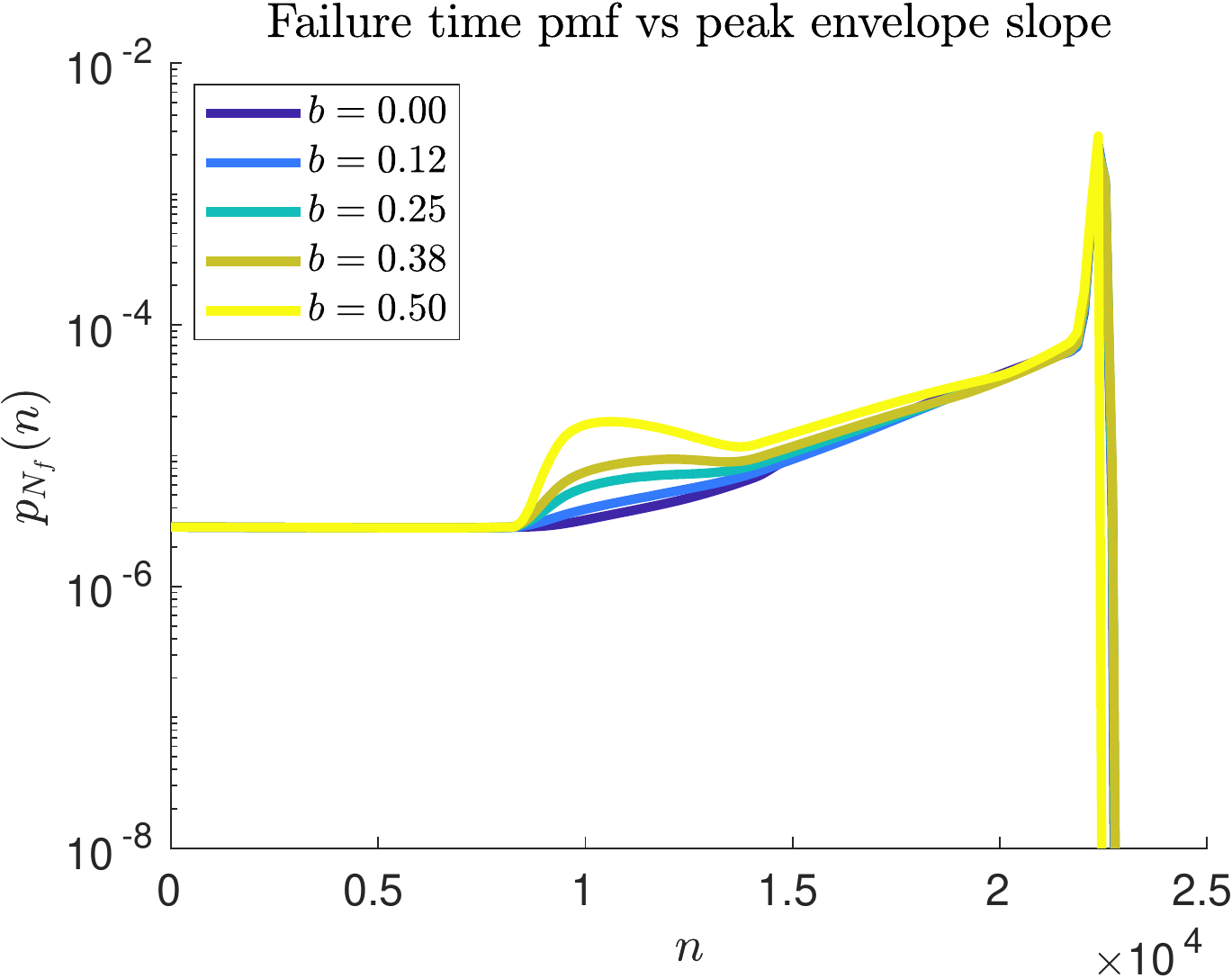}
\includegraphics[width=0.45\linewidth]{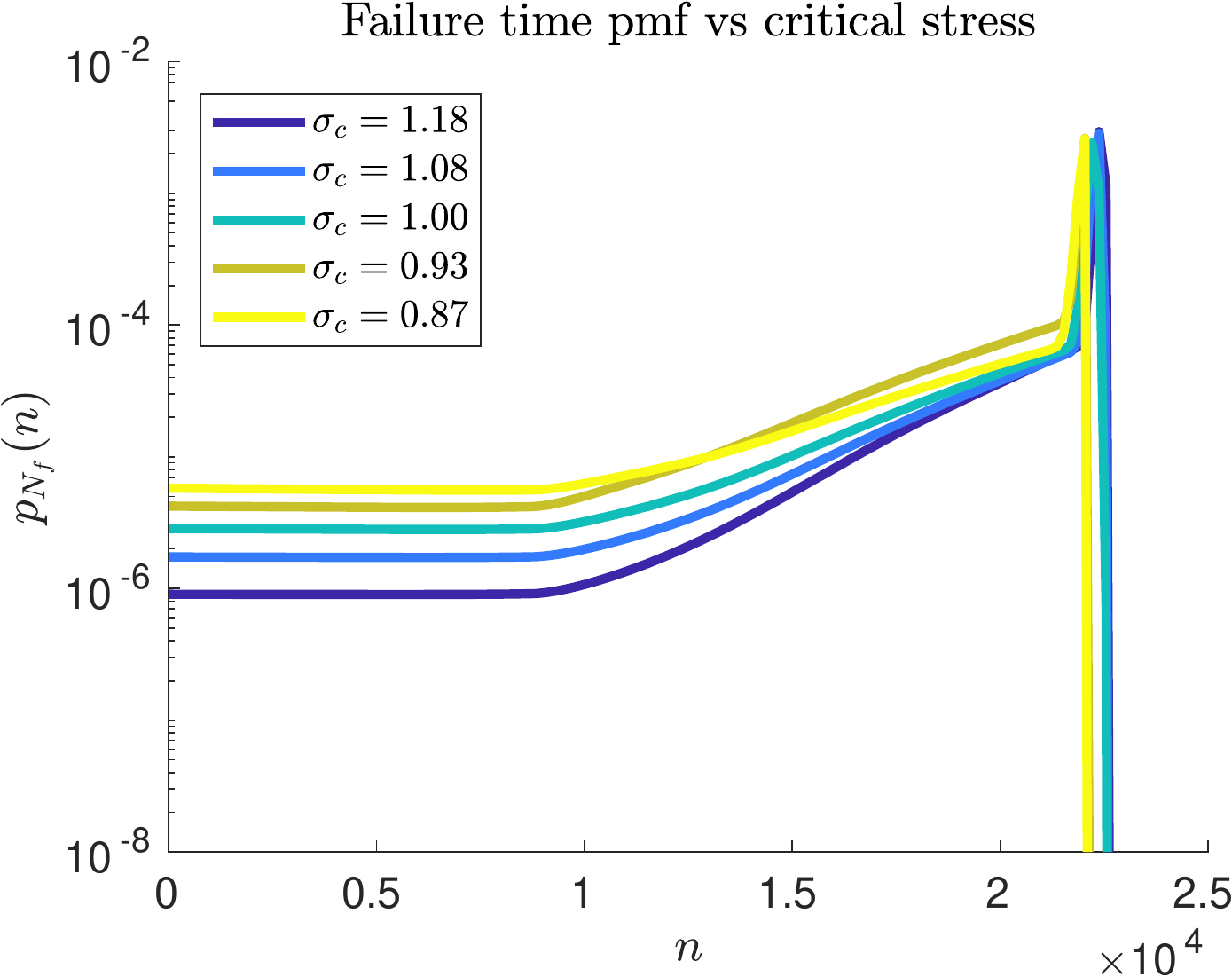}
\includegraphics[width=0.45\linewidth]{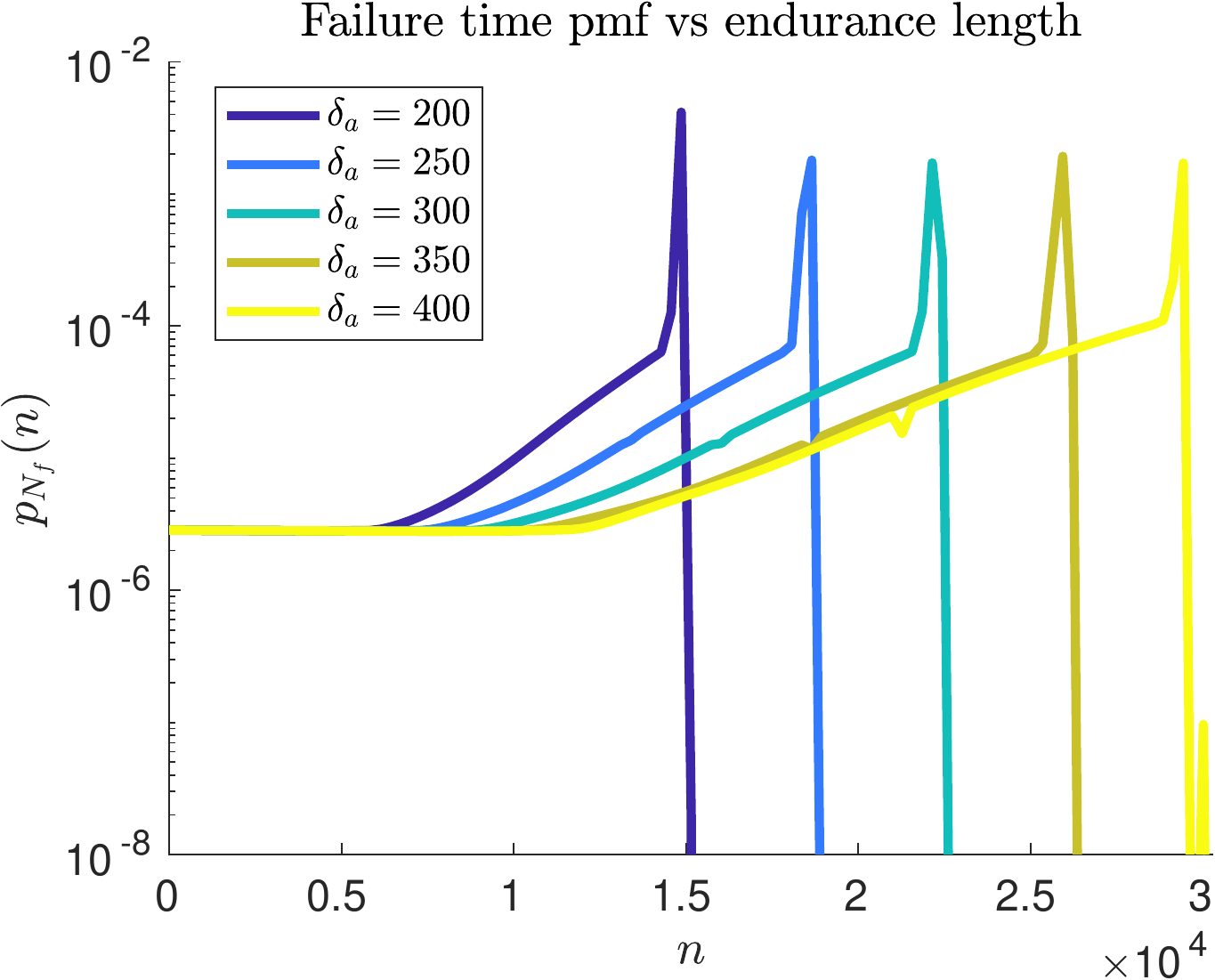}
\caption{Comparison of the failure time distributions for different parametric
choices of the coherent envelope.  a)  Initial slope, which sets value of
$K_0^+$.  b)  Peak slope, which affects the shape of $\kappa(\sigma)$ and
$\eta(\sigma)$ near $\sigma \approx \sigma_c$.  c)  Critical stress $\sigma_c$,
which determines what magnitude load leads to immediate failure.  d)  Endurance
length $\delta_a$, which directly affects the typical lifetime.}
\label{fig:spline-comparison}
\end{figure}

Figure \ref{fig:spline-comparison} compares the failure time distribution of different coherent envelopes.  Subplots (a) and (b) show the dependence on the two adjustable parameters of the spline.  Adjusting the initial slope and peak slope have only minor effects of the shape of the distribution.  Specifically, adjusting the initial slope, i.e. $K_0^+$ has as a result the expected shift of the peak of the distribution. On the other hand, modifying the value of $b$ has only negligible effects. Moreover, adjusting $\sigma_c$ (subfigure $c$) has, as expected, a very significant effect on the probability of early failure, without significantly effecting the median failure time.  Similarly, adjusting $\delta_a$ has a large effect on the median failure time without significant effects on the extremely early failure time. Therefore, we conclude that the exact shape of the coherent envelope is not critical for the general form of the derived pmf for the failure time (e.g. form of the tail). However, it is essential to know, beyond the standard parameters $\sigma_c$, $\delta_a$, and the $SN$ curve, the initial slope of the coherent envelope (inherently connected with $K_0^+$).

\section{Conclusions}


We have considered the SO model for the probabilistic characterization of fatigue lifetime in the case of random stochastic loads with intermittent characteristics. We first developed an efficient numerical approximation  scheme for the direct numerical simulation of a given load time series. Next we developed analytical approximations for the probability mass function of the failure time, which takes as inputs the coherent envelope and the statistical characteristics of the loads, including their non-Gaussian intermittent features.  These analytical expressions are practical and allow for the substitution of time consuming Monte
Carlo simulations, especially important in the high-dimensional spaces associated
with structural finite element models. Together with our analysis we provide open-source codes that implement the derived formulas for general pmf of the loads.

Using the analytical approximation of the failure time probability distribution, we showed that the SO model predicts a much wider spread in failure times than the Palmgren-Miner rule for materials subject to intermittent loading, a spread we expect to better match realistic fatigue lifetimes.  In particular, the SO model identifies a long left tail of early failure times, and identifies an association between certain early loads and subsequent early failures--the ascending leg intersections.

Finally, we have examined the robustness of the derived probability mass functions for failure time with respect to the coherent envelope geometry. In particular, we have shown that with constrained $SN-$curve the resulted probability distribution has  practically invariant shape, an important property so that existing experimental data can be utilized for the presented analysis.  





\subsubsection*{Acknowledgments} We are grateful to Prof. Michael Ortiz who introduced us the SO model. We are also thankful to Prof. Vladas Pipiras for useful advice regarding the probabilistic treatment of the $argmin$ functions. This research has been supported by the ONR Grant N00014-20-1-2366, a Doherty Career Development Chair and a Mathworks Faculty Research Innovation Fellowship. 

\subsection*{Code} All the codes are available in the link: \texttt{https://github.com/batsteve/terrifying-eruption}

\bibliographystyle{abbrv}
\bibliography{bib_file.bib}

\end{document}